\documentclass[final,8pt,5p,times,twocolumn]{elsarticle}

\usepackage{amssymb}
\usepackage{caption}
\usepackage{comment}
\usepackage{listings}
\usepackage{amsmath}
\usepackage{subcaption}
\usepackage[dvipsnames]{xcolor}
\usepackage{algorithm,algorithmic}
\usepackage[hyphens]{url}
\usepackage[hidelinks]{hyperref}
\usepackage{spverbatim}
\hypersetup{breaklinks=true}

\newcommand{\framework}{Ripple}
\newcommand{\frameworkSpace}{Ripple }
\newcommand{\rsec}[1]{Section~\ref{#1}}

\newcommand{\rlist}[1]{Listing~\ref{#1}}
\newcommand{\rfig}[1]{Figure~\ref{#1}}
\newcommand{\rtab}[1]{Table~\ref{#1}}
\newcommand{\ralg}[1]{Algorithm~\ref{#1}}

\lstdefinestyle{cpp}{
    language=C++,
    basicstyle=\scriptsize\ttfamily,
    captionpos=b,
    frame=tbrl, % draw a frame at the top and bottom of the code block
    tabsize=2, % tab space width
    showstringspaces=false, % don't mark spaces in strings
    commentstyle=\color{ForestGreen}, % comment color
    backgroundcolor=\color{Gray!5},
    keywordstyle=\color{blue}, % keyword color
    stringstyle=\color{red}, % string color,
    morekeywords={constexpr, size_t, emplace, then, split, then_split, reduce, then_reduce, concurrent_padded_access, exclusive_padded_access},
    aboveskip=\smallskipamount,
    belowskip=\smallskipamount
}

\journal{Journal of Parallel and Distributed Computing}

\begin{document}

\begin{frontmatter}



\title{Ripple : Simplified Large-Scale Computation on Heterogeneous Architectures with Polymorphic Data Layout}


\author{Robert Clucas \corref{cor1}}
\ead{rjc201@cam.ac.uk}

\author{Philip Blakely}
\ead{pmb39@cam.ac.uk}

\author{Nikolaos Nikiforakis}
\ead{nn10005@cam.ac.uk}

\address{
Maxwell Centre,
Cavendish Laboratory,
JJ Thomson Avenue,
Cambridge,
CB3 0HE}

\cortext[cor1]{Corresponding author}

\begin{abstract}

GPUs are now used for a wide range of problems within HPC. However, making efficient use of 
the computational power available with multiple GPUs is challenging. The main challenges in 
achieving good performance are memory layout, affecting memory bandwidth, effective 
use of the memory spaces with a GPU, inter-GPU communication, and synchronization. We address
these problems with the Ripple library, which provides a unified view of the computational space 
across multiple dimensions and multiple GPUs, allows polymorphic data layout, and provides a simple 
graph interface to describe an algorithm from which inter-GPU data transfers can be optimally scheduled. 
We describe the abstractions provided by Ripple to allow complex computations to be described
simply, and to execute efficiently across many GPUs with minimal overhead. We show performance results for a
number of examples, from particle motion to finite-volume methods and the eikonal equation, as well as showing
good strong and weak scaling results across multiple GPUs.

\end{abstract}




\end{frontmatter}

\section{Introduction}\label{sec:introduction}

The heterogeneous nature of compute servers containing multiple many-core CPUs as well as
a large number of accelerators---usually GPUs---makes it difficult to take full advantage of the available computational resources. Programming one domain (i.e.~CPU or GPU) effectively requires expert knowledge of the respective hardware and associated tools, but is especially true for GPU implementations,
where the programming model is significantly different from that required to 
program many-core CPU-based systems. It is also typically more difficult to extract high levels
of performance from accelerators due to the difference in hardware design between CPUs and GPUs, 
which makes the performance of GPU code very dependent on register use, lack of indirection, and regular 
memory access patterns. These effects are magnified once multiple
accelerators are present in a system due to cost of data transfer between them.

Most existing software frameworks separate the computational 
domains, or provide improvements in either programmer productivity or performance, but not both. 
Therefore, we identify two categories which need to be addressed to achieve 
good performance on systems with many CPU cores as well as multiple accelerators: fine-grained 
parallelism and coarse-grained parallelism.

Within fine-grained parallelism, there is the parallelisation of data and the execution of the 
computation on the data. These are usually the kernels which run on the accelerators,
or across the many CPU cores in the system. Achieving good performance in this area requires tuning 
the data layout, kernel execution sizes, and use of specific memory regions on the computational 
devices (i.e. cache and shared memory).

Coarse-grain parallelism involves the scheduling
of kernels which execute on the computational resources, partitioning of work across the heterogeneous
devices, and the reduction of latency when data must be transferred or shared between the different
devices.

There is very little existing work which addresses both areas, and specifically all mentioned sections
within each area, effectively. Here we present \framework \footnote{\url{https://robclu.github.io/ripple_docs/}}, a software library which 
addresses the above problems, providing an interface for specifying both coarse- and fine-grained 
parallelism, as well as providing the user with a simple means for configuring all aspects of parallelism to allow good performance to be achieved for their application with minimal effort, 
and with limited knowledge of GPUs or parallel and distributed systems. The major contributions 
of our work are the following:

\begin{itemize}
    \item \textbf{Polymorphic data layout for user-defined types:} Users can specify data layouts
    for user-defined classes as a template parameter, which allows the class to be used as either 
    Array-of-Struct (AoS, contiguous layout) or Struct-of-Array (SoA, strided layout). We 
    provide interoperability of the layouts so that they can be used seamlessly throughout an application.
    This allows users to determine the best layout for the problem, and to use different data layouts for 
    each part of an application or execution domain.
    \item\textbf{Specification of the memory space:}
    The user can specify which memory space should be used 
    for each block of the data involved in a computational kernel (for example, shared or global memory on the GPU). 
    Achieving good performance on GPUs requires testing all memory spaces and data layouts for a problem, which is cumbersome for large applications. 
    We allow the memory space to be specified (in a single line) at the point of computation, allowing the same data to be used 
    in different memory spaces for different parts of an algorithm.
    \item \textbf{N-dimensional tensor data type and iterators:} We provide 
    an N-dimensional tensor data type and an iterator interface over the space. This removes the need for error-prone manual memory allocation 
    and index computations. Further,
    padding cells can be specified for the tensor, required for many stencil-type operations, and 
    allow these can be filled from neighbouring tensors where necessary.
    \item \textbf{Expressive graph model for the structure of a series of kernels:}
    Kernels, and the implicit and explicit dependencies between them, can be specified at a high level as 
    a Directed Acyclic Graph (DAG). The memory and computational dependencies can then be determined by the 
    library allowing optimisation of kernel-scheduling to minimise the overall latency of the computation.
    \item \textbf{Heterogeneous computation} Kernel-execution can be specified in a graph for execution on the CPU or GPU, enabling heterogeneous computation and making it possible to fully
    utilise the available computational power of both domains in a system.
\end{itemize}

While our library allows general computation to execute on heterogeneous nodes, our focus is on computations performed on multi-dimensional tensor-like data partitioned across multiple GPU accelerators. We take this approach due to modern GPUs providing significantly 
more computational performance than CPUs, and their ability to process regularly laid-out data efficiently. 
Despite their much higher theoretical computational throughput, attaining this on the GPU is more difficult than
for the CPU, so simplifying programmer effort for better utilisation of GPU accelerators has the largest overall 
benefit---particularly for problems which require non-trivial data structures. 

\frameworkSpace requires a compiler implementing the C++-17 standard and therefore supports Clang 9, 10, and 11, with CUDA \cite{Cuda2020} versions 9.0 
though 11.2, as the compiler for both host and device code, or CUDA versions $>=$ 11 with {\tt nvcc} as the device 
compiler with GCC versions $\ge$ 9.3.0 or Clang versions $\ge$ 9.0.0 as the host compiler, since CUDA 11.0 provided the first nvcc compiler
with C++-17 support for device code.

The remainder of the paper is structured as follows:
In \rsec{sec:related_work} we review existing work. \rsec{sec:abstraction}
gives an overview of the abstractions in \framework. These are described in more detail in \rsec{sec:fine_abstractions} where we describe the n-dimensional tensor data type,
polymorphic data layout functionality, a simplified
method for accessing elements in polymorphic data types, and the iterator abstraction
which allows efficient accessing of tensor data in kernels. At a higher level, \rsec{sec:coarse_abstractions} outlines heterogeneous allocators in \frameworkSpace
for improving the performance of workflows with dynamic memory requirements, the graph interface
for computational workflow description, how constraints can be placed on kernel data 
access to specify data dependencies between kernels in a graph, and how to schedule work 
between the CPU and GPU computational domains. In \rsec{sec:scheduling} we describe how work is scheduled on either CPU or GPU as appropriate. Then, in \rsec{sec:data_layout_results} we provide benchmarks for the performance of 
the fine-grained aspects of our library, while \rsec{sec:scaling_results} shows the scaling performance of the library across
multiple computational devices. In \rsec{sec:future_work} we discuss the main limitations of the library and directions of 
future work, and \rsec{sec:conclusion} summarises the whole work.

%=================================================================
% RELATED WORK
%=================================================================
\section{Related Work}\label{sec:related_work}

While most related work falls into either the fine- or coarse-grained categories, SYCL \cite{sycl} does provide features
for both. The latest revision has support for kernels written in C++17, which can be executed on a number of backends, allowing
a lot of flexibility for supported hardware. The framework also allows for kernels to be executed across
multiple devices if available. However, a lot of the work has to be done manually by the user, such as the partitioning of
data and transfer of data between execution devices. The main limitations of SYCL are a lack of support for CUDA devices as
a backend, which likely results in less efficiency on CUDA supported devices as OpenCL does not provide all the features available
in CUDA. Additionally, a number of the abstractions we provide to make programming simpler and more
efficient in a multi-GPU context, such as polymorphic data layout, iterators, multi-dimensional tensors, and the graph interface 
which handles data transfer between devices, are not available in SYCL.

\subsection{Fine-grained Parallelism}

There are a number of existing frameworks which allow code to be parallelised by the compiler
following programmer directives. Examples
are OpenMP \cite{openmp} and OpenACC \cite{openacc}. They are simple to use, but are
limited in the complexity of computation which they can parallelize effectively. OmpSs \cite{openmps} extends 
OpenMP to allow heterogeneous computation, but still requires a large amount of work from the programmer 
and does not provide cross-GPU functionality. 
These perform very well at parallelising simple loops, but are not able to handle more 
complex computations. A good overview of the benefits and limitations of these types of programming models 
is provided by Lee and Vetter \cite{Lee2012}.

Kokkos \cite{EDWARDS2014} is a more sophisticated solution which has seen
significant use, particularly in scientific applications. 
It provides data-layout flexibility for simple data-types (not user-defined), and high-level 
constructs for offloading computation to one of the supported many-core backends. While the functionality is vast, 
and performance is very good relative to raw CUDA, the interface is quite involved, particularly for custom data structures. The main limitation of Kokkos is that it requires the programmer to manage 
inter-GPU communication and data transfer explicitly with another library, such as MPI, thus
requiring a good knowledge of the underlying hardware and its API. Additionally, it does not 
provide an interface for shared memory use on the GPU or polymorphic data layout, which can both provide significant 
performance improvements for some applications, as we show later.

\subsection{Coarse-grained Parallelism}

Task based programming models have recently become popular due to their ability to specify complex computational
workflows at a high level, and have them run efficiently in parallel. These are particularly suited to CPUs due to the dynamic and 
irregular nature of the tasks in the computational flow. Early works are StarPU 
\cite{Augonnet2011}, which provides a scheduling environment for tasks to run on heterogeneous architectures, 
exhibiting good performance. HPX \cite{kaiser2014} is similar, but is C++ standard conforming and allows task-based 
work to be executed on larger-scale heterogeneous architectures. It provides similar functionality to our graph
interface for specifying a computational flow, but does not provide fine-grained data-space
abstractions or polymorphic data-layout, which reduce development complexity
and improve performance. A more recent work is Cpp-Taskflow v2 \cite{huang2020}, 
which has seen significant use of late. The second version extends the original implementation to allow tasks 
to be executed heterogeneously on both the CPU and GPU, but does not provide any GPU based abstractions for data spaces or data transfer between GPUs.

While all of these works provide good abstractions for task-based parallelism, they offer limited support for the fine-grained parallelism which is required 
to take advantage of the throughput provided by GPU accelerators.

Once multiple accelerators are present in a system, most applications require data-transfer 
between the accelerators which contributes significantly to the overall computation time due to the bandwidth 
and latency of the memory system, resulting in under-utilization of the accelerators as they wait for data to process.

Work addressing these issues includes CuMAS \cite{Belviranli2016}, which looks at reducing the latency of data-transfer with respect to GPUs, but for the multi-application space, 
rather than within a single specific application. Bastem {\it et al.}~\cite{Bastem2017} 
provide a framework for overlapping data transfers and compute on the GPU, however, the overall use is limited by the requirement 
for kernels to be written using OpenACC, which they show to produce sub-optimal performance compared to CUDA. Lastly, Daino \cite{Wahib2016} is a framework for hiding memory-transfer latency, but
specifically for Adaptive Mesh Refinement, which limits its scope.

%=================================================================
% ABSTRACTION
%=================================================================
\section{Abstractions}\label{sec:abstraction}

Modern computational environments contain multiple nodes which communicate via an interconnect. Ideally the system architecture would be abstracted by the programming interface, but this is 
challenging, especially when the data transfers create
dependencies between operations on different devices in the system. Thus
there is a separation into coarse- and fine-grained parallelism, for which we provide a number 
of abstractions to allow the programmer to maximise performance in both areas.

The data abstractions we provide are tensors, polymorphic data layout, accessors, and allocators, which target fine-grained
parallelism, while the execution abstractions, which facilitate coarse-grained parallelism, are iterators and graphs. We describe these in more detail in the following sections.

\section{Data Abstractions}\label{sec:fine_abstractions}
Here we describe the abstractions that deal with data; generally improving fine-grained parallelism.
%=================================================================
% TENSORS
%=================================================================
\subsection{Tensors}\label{sec:tensors}

%--------------------
% SOA/AOS figure
%--------------------
\begin{figure*}[t]
    \centering
    \begin{subfigure}{.48\linewidth}
    \includegraphics[width=\columnwidth]{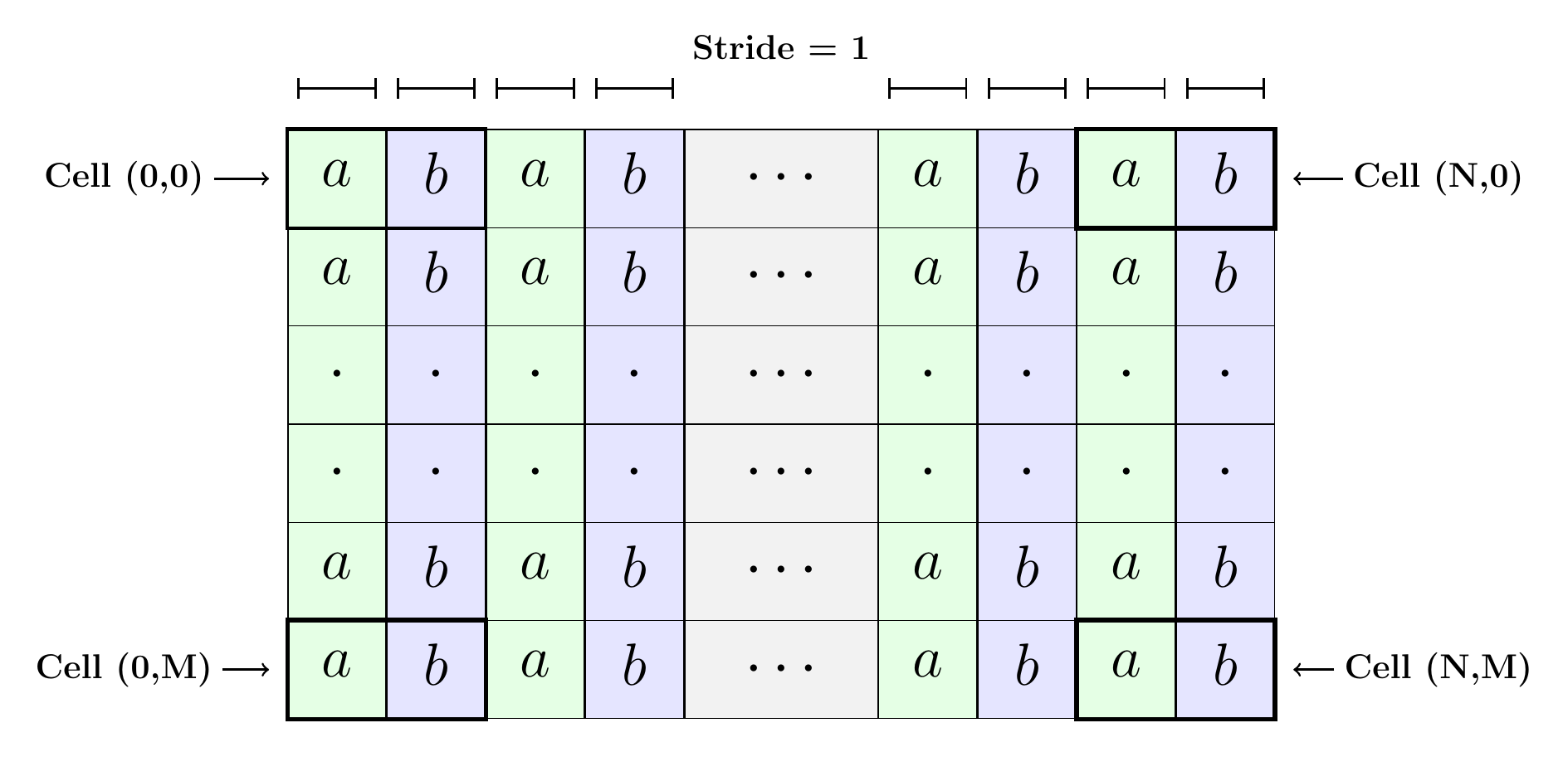}
    \caption{Contiguous layout.}
    \label{fig:contig_layout}
    \end{subfigure}
    \hfill
    \begin{subfigure}{.48\linewidth}
    %\label{fig:strided_layout}
    \includegraphics[width=\columnwidth]{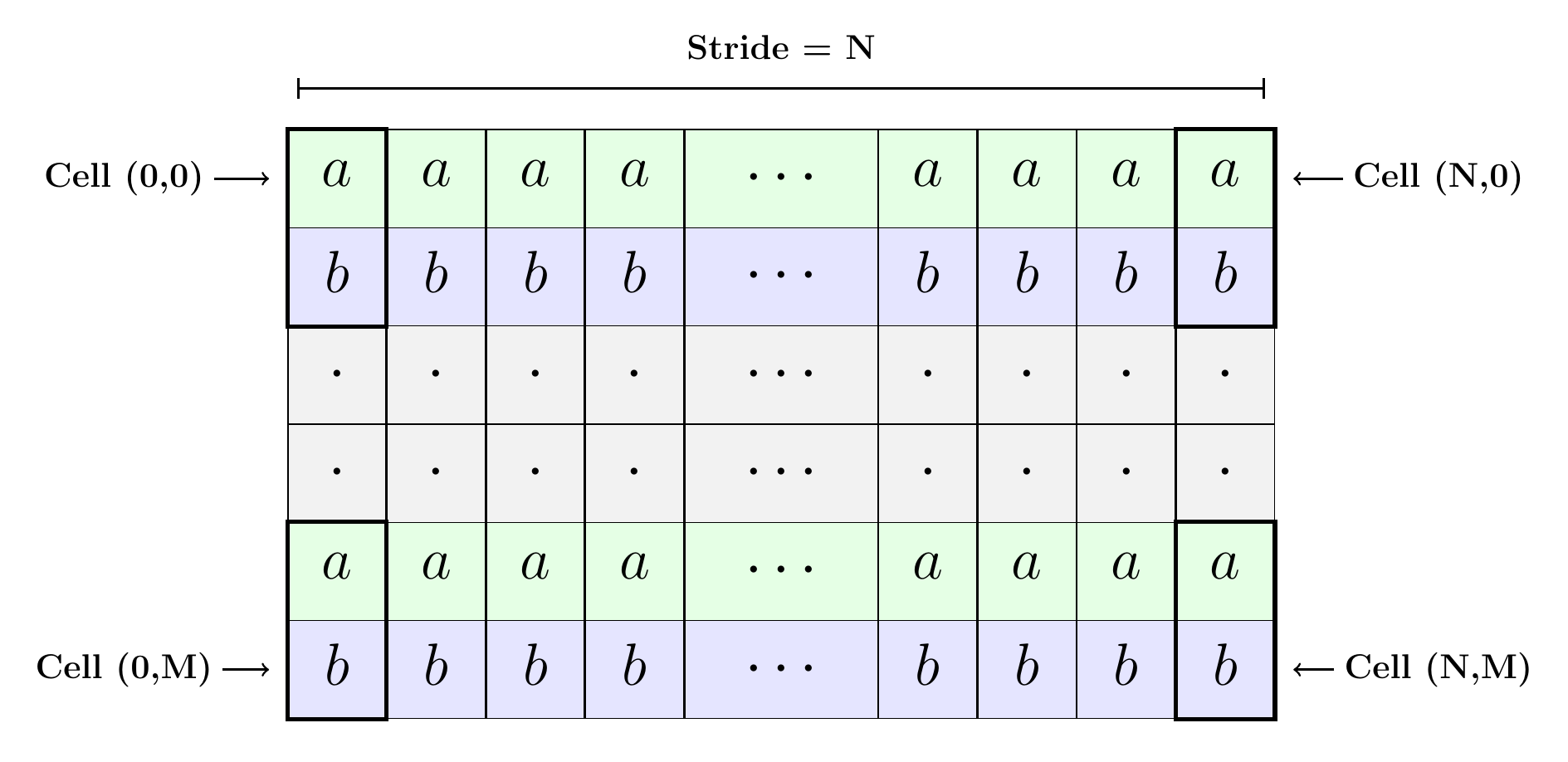}
    \caption{Strided layout.}
    \label{fig:strided_layout}
    \end{subfigure}
    \caption{Illustration of strided and contiguous layouts for a struct with two components, $\rho$ and $p$.}
    \label{fig:poly_layout}
\end{figure*}

In \framework, a tensor is an abstraction over a multi-dimensional space. It contains a data type, which
is stored for each point in the multi-dimensional space, and the dimensionality of the space, both provided at compile-time as template parameters. Data types can be scalar, such as 
\verb|double,float,int|, etc.~or user-defined. When the type is user-defined, the data-layout data can be specified as polymorphic, allowing the storage of the class data in the tensor 
to be SoA (elements are strided) or AoS (elements are contiguous), as described in 
\rsec{sec:layout_flex}, and shown in \rfig{fig:poly_layout}. When initializing the tensor data, the layout is 
inferred from the data type, and the size of each dimension in the space is specified as a run-time parameter 
when the tensor is created, which can be resized if necessary.

Many operations require the use of data within a stencil. For example, an approximation 
of the first derivative of f using a forward difference in the x dimension at index u is
\begin{equation}
    \Delta_x f_u = f_{u + 1} - f_u
\end{equation}
which requires access to the data point $f_{u + 1}$. If $f_u$ is at the boundary of the domain, the data at 
$f_{u + 1}$ is invalid, so padding data is required, which can be specified as a parameter 
when a tensor is created. \frameworkSpace provides a number 
of methods for loading the padding for common cases---such as first order
and constant extrapolation. Padding becomes more complicated when the tensor data is strided or exists 
on multiple devices, potentially introducing a memory dependency in the DAG
representing the computation, and \frameworkSpace simplifies the handling of the padding data in such cases.
Through the graph interface, options are provided for the strictness of the dependency on operations which use or modify
padding data, allowing the computational graph to be optimized to minimise the latency of memory operations as well as removing
the need for the programmer to implement these themselves. These operations are described fully in \rsec{sec:graph}.

The tensor data can be partitioned across the system devices, and also within each device. An example is shown in \rlist{lst:tensor_split}. This only requires changing a single parameter, making bench-marking straightforward. Partitioning across multiple dimensions results in smaller blocks in each of the partitioned
dimensions. However, this requires data to be transferred between the devices if neighbour data 
is accessed, introducing dependencies in the DAG. We suggest partitioning across a single dimension for best 
performance. For multi-dimensional tensors, partitioning the higher dimensions is likely the best strategy
because it results in a memory transfer of data which has a coalesced access pattern (i.e.~accesses contiguously laid out
data).

As shown in \rlist{lst:tensor_split}, the tensor interface allows both the number of partitions per dimension, as 
well as the number of sub-partitions to be specified for each dimension. Sub-partitions 
exist on the same device so may introduce a memory dependency where one might not be necessary. However, they may also increase 
the overall opportunity for parallelism in an application, or create partition sizes which are suitable for both the CPU and GPU,
leading to improved performance.

%=================================================================
% DATA LAYOUT FLEXIBILITY
%=================================================================
\subsection{Polymorphic Data Layout}\label{sec:layout_flex}

Tensor data layout is determined by the data type, which is always contiguous, unless the 
data type is user-defined and inherits from the \verb|PolymorphicLayout| struct, which requires
a template parameter to specify the type of layout. This allows user-defined types to have flexible data layout, 
and therefore the option to profile 
which layout is the best for different parts of an application. In \framework, the layout of the class 
data can be defined as \textit{contiguous} or \textit{strided}. Contiguous data has the elements of the struct 
laid out contiguously in memory, while strided data is laid out with a stride equal to the size of the 
first dimension in the tensor. Internally, strided data stores an array of each of the
components of the struct contiguously, with the component arrays being stored successively. These two 
layouts are shown in \rfig{fig:poly_layout}.

Creating a polymorphic data layout requires a user-defined class to specify the types stored using the 
\verb|StorageDescriptor| template class, which takes the layout and a list of types to be stored, similar to 
\verb|std::tuple|. We provide a \verb|Vector<T, Size>| class to specify that \verb|Size| 
elements of type \verb|T| should be stored in a vector-like container. 
There are existing solutions which allow polymorphic layout within containers, but only for 
built-in types such as \verb|double, float, int|, etc. Additionally, accessing the elements is done
through an index, which does not provide the context that the name 
of a struct member does, and does not allow the polymorphic data to be used as both an object and within the tensor. 
Both of these limitations are significant in the development of large software projects.
\frameworkSpace allows heterogeneous types to be stored in the tensor, 
and class-data to be accessed through the \verb|StorageDescriptor| as if it were a 
\verb|std::tuple|. The types must be defined through the \verb|StorageDescriptor| 
interface rather than directly within in the class, and to inherit from the \verb|PolymorphicLayout<T>| trait-class, 
allowing \frameworkSpace to determine that the user-defined type has a polymorphic memory layout with no 
runtime-overhead. The \verb|StorageDescriptor|
provides an alias for the actual storage type, and the elements can be accessed though 
the \verb|get| template method which computes offsets at compile time. An example is shown in 
\rlist{lst:layout_example} which defines a state class with \verb|D| elements of type \verb|T| and a \verb|bool| 
indicating whether the state is valid. When used with a tensor, this will lay out the data based on 
the type provided as the \verb|Layout| template parameter.

\begin{lstlisting}[style=cpp,float=t,caption=Tensor domain split example. Solid lines are device boundaries.,label={lst:tensor_split}]
/* Create a 2D tensor of doubles:
 *  - 2 partitions per GPU in dimension 0 (x)
 *  - 2 partitions per GPU in dimension 1 (x)
 * The resulting data sizes across the devices is:
 *   (500, 500) | (500, 500)
 *   -----------------------
 *   (500, 500) | (500, 500)
 */
constexpr size_t size_x = 1000;
constexpr size_t size_y = 1000;
Tensor<double, 2> t({2, 2}, size_x, size_y);

/* Create a 2D tensor of doubles:
 *  - 2 paritions dimension 0 (x)
 *  - 2 sub-partitions per partition in dimension 0 (x)
 *  - 2 paritions per dimension 1(y)
 *  - 2 sub-partitions per partition in dimension 1 (y)
 * The resulting data sizes across the devices is:
 *   (250, 250) (250, 250) | (250, 250) (250, 250) 
 *   (250, 250) (250, 250) | (250, 250) (250, 250)
 *   ---------------------------------------------
 *   (250, 250) (250, 250) | (250, 250) (250, 250)
 *   (250, 250) (250, 250) | (250, 250) (250, 250)
 */
constexpr size_t size_x = 1000;
constexpr size_t size_y = 1000;
Tensor<double, 2> t({2, 2}, {2, 2}, size_x, size_y);
\end{lstlisting}

\begin{lstlisting}[style=cpp,float=t,caption=Polymorphic layout example,label={lst:layout_example}]
template <typename T, size_t D, typename Layout>
struct State : 
  public PolymorphicLayout<State<T, D, Layout>> {
  // Declare storage for D elements of type T and a bool.
  using Desc    = StorageDescriptor<
    Layout, Vector<T, D>, bool>;
  using Storage = typename Desc::Storage;
    
  // Normal access for vector:
  //  Gets the ith component from the vector
  auto operator[](size_t i) -> T& {0
    return storage.get<0>(i);
  }
    
  // Density is element 0 of type 0 from storage:
  auto density() -> T& {
    return storage.get<0, 0>();
  }
  // Pressure is element 1 of type 0 from storage:
  auto pressure() -> T& {
    return storage.get<0, 1>();
  }
  auto is_valid() const -> bool {
    return storage.get<1>();
  }

  Storage storage;
};

// Specialization of the type for strided:
using State3d = State<double, 3, StridedView>;
\end{lstlisting}

%=================================================================
% Accessors
%=================================================================
\subsection{Accessors}\label{sec:accessors}
\begin{lstlisting}[style=cpp,float=t,caption=Vec implementation example using accessors,label={lst:vec_accessor}]
template <typename T, size_t D, typename L>
class Vec : public PolymorphicLayout<Vec<T, D, L>> {
  using Desc    = StorageDescriptor<L, Vector<T, D>>;
  using Storage = typename Desc::Storage;
  
  // Accessors types:
  using X = Accessor<T, Storage, 0>;
  using Y = Accessor<T, Storage, 1>;
  using Z = Accessor<T, Storage, 2>;
  using W = Accessor<T, Storage, 3>;
    
  union {
    Storage storage;
    X x; Y y; Z z; W w;
  };
};

// Usage
Vec<int, 3, ripple::ContiguousOwned> v;
v.x = 10, v.y = 20;
\end{lstlisting}
For some use cases, it is more descriptive to define the members of a class as public, rather than
as member functions which access the associated components from the storage, as shown in 
\rlist{lst:layout_example}. One such example is that of a vector class, with components
for \verb|x,y,z|, etc., which could be implemented
with \verb|x,y,z| as public members that can be accessed directly. However, if the storage is 
a contiguous array, they would have to be defined as public member functions for getting and setting
the values, or would have to be stored as public members, which would require specialization for each number of elements.

\begin{comment}
The MSVC compiler has functionality to support replacing 
function syntax for getting and setting members with \verb|__declspec(property(...))|, however, it 
is not portable and does not support the polymorphic layout functionality.
\end{comment}
We 
provide an \spverb|Accessor<ValueType, StorageType, Index>| abstraction to allow a given index in the 
storage to return a given value type, while still allowing the polymorphic layout flexibility. The 
accessor(s) can be used in a union with the storage type for a class to provide a more concise interface for data access. 
Additionally, nested unions can be used to overload the functionality of 
each of the storage elements, allowing access as \verb|x| 
or \verb|r|, depending on context. An example is shown in \rlist{lst:vec_accessor} for a 
vector class. The advantage of the accessor implementation over a public member implementation
is that the class data can be laid out as strided or as contiguous data when used with the
tensor.

\section{Execution abstractions}\label{sec:coarse_abstractions}
In this section we describe abstractions relating to the flow control; typically associated with coarse-grained parallelism.

%=================================================================
% ALLOCATORS
%=================================================================
\subsection{Allocators}\label{sec:allocators}

Often memory requirements are not known upfront, and therefore require memory to
be allocated dynamically. However, this significantly reduces performance
since \verb|cudaMalloc| and \verb|cudaFree| require synchronization to ensure 
that the memory being allocated or freed is not touched before or after the operation. 
These calls introduce significant points in the execution pipeline of an application where no work
is being performed on either the CPU or the GPU. It is thus important to avoid dynamic allocation during a
computation. For the Euler example presented in \rsec{sec:scaling_results}, dynamic allocations, even 
accounting for only 2\% of the total memory requirement, caused performance reductions of up
to 30\%.

We therefore provide an
allocator, \verb|MultiarchAllocator|, which provides dynamic memory allocators for the CPU and the GPU via the \verb|cpu_allocator()| and \verb|gpu_allocator()| methods, respectively. For the 
CPU, we allocate a buffer of pinned host memory, since this is faster for transfer to the GPU and 
is required for asynchronous copies between the CPU and GPU. For the GPU we allocate a buffer with 
\verb|cudaMalloc|, from which the allocator allocates. We use the same allocation strategy for the CPU 
and GPU allocators, which is to allocate linearly from the buffer. 
The CPU allocator has a buffer per thread, while the GPU allocator has a buffer per GPU, and both
are thread safe to allow them to function correctly with the scheduler described in \rsec{sec:scheduling}. 
Since the linear allocation strategy is relatively simple, we plan to extend the allocation strategy in future. For our 
use cases, however, the current implementation significantly improved overall performance in multiple
applications. This was also the case for performing an out of place reduction across multiple devices, 
since the results buffers for the the reduction of the individual devices did not need to be allocated
reducing the time for which the GPU was idle. We also provide constructors for the 
tensor data type which allow it to take a \verb|MultiarchAllocator| from which it can allocate, for 
tensors requiring dynamic creation or resizing. 

%=================================================================
% ITERATORS
%=================================================================
\subsection{Iterators}\label{sec:iterators}

When a callable is defined and operates on a tensor, its parameter type
for each tensor is either an \verb|IndexedIterator<T, Space>| or a \verb|BlockIterator<T, Space>|, 
where \verb|T| is the type of the tensor data and \verb|Space| is the multi-dimensional space of the 
block within the tensor on which the iterator can iterate. The layout of the data iterated over is always the same as that of the tensor, and includes padding data if present. The iterator is automatically offset to the relevant cell in the tensor, avoiding the need 
for the user to have to perform any offsetting over padding data. The layout of the data is hidden by the 
iterator so that it can iterate over both contiguous and strided data.

\subsubsection{Indexed Iterator}

\begin{lstlisting}[style=cpp,float=t,caption=Iterator offsetting in multiple dimensions,label={lst:iter_offset}]
/* Creates a callable which takes an iterator, computes 
 * the central difference in each dimension, and sums it.
 * We use auto for the iterator type for brevity.
 */
auto callable = [] ripple_host_device (const auto& it) {
  double diff_sum = 0.0;
  unrolled_for<iterator_traits_t<Iterator>::dims>(
    [&] (auto dim) {
      diff_sum += *it.offset(dim,  1) 
                - *it.offset(dim, -1);
  });
};
\end{lstlisting}

The \verb|IndexedIterator| carries with it indices which define the location of
its block within the tensor, as well as its indices in the global iteration space, allowing 
the iterator to provide thread, block, and global indices and sizes. This allows the multi-dimensional space, partitioned across multiple devices, to appear
as a single space. The iterator, however, can only iterate within a single block in the partitioned tensor 
space, due to the tensor data for different blocks potentially being on different devices.

This has an overhead per thread of 12 bytes per dimension 
in the tensor and iterator. However, benchmarks have shown the performance effect to be small, and the increase in 
ease of use of iterators compared to raw multi-dimensional indices is significant. The performance 
implications of the iterator framework are provided 
in \rsec{sec:data_layout_results}. An additional benefit of the iterator is that the syntax for accessing
data is consistent for different dimensions. Most frameworks overload \verb|operator()| to enable indexing in multiple dimensions requiring separate specializations for each number of dimensions. While sometimes necessary,
other operations differ only in the number of dimensions, thus, the iterator provides a 
compile time trait for the number of dimensions, and an \verb|unrolled_for<dims>(operation)| method is available 
to unroll an operation for a number of dimensions as shown in \rlist{lst:iter_offset}. We have 
benchmarked code which uses \verb|unrolled_for| against a regular loop with a runtime value and have 
observed slightly better performance, and lower register usage on the GPU, in almost all cases.

The iterator is similar to that in the standard C++ library, with extensions to
support a multi-dimensional space. An iterator can be offset in a given dimension, as shown in \rlist{lst:iter_offset}, 
and de-referencing provides access to the underlying offset data. Calls to the 
offset method can be chained to offset in multiple dimensions. When a callable is executed, a check is performed
on the iterator's index against the the bounds 
of the multi-dimensional iteration space so that any out of range iterators do not execute the callable, ensuring that 
iterators only run on data valid within the tensor space.

\subsubsection{Block Iterator}

A \verb|BlockIterator| is a
lightweight version of the \verb|IndexedIterator|, and used in more performance critical contexts. 
It iterates over a subspace of a tensor partition, and does not 
contain index information for the global space so cannot determine its global location. It can, 
however, provide the index with the subspace it can iterate over. A
\verb|BlockIterator| can be created over both statically and dynamically sized spaces. In the statically sized case, 
the size of the iterator is the size of a pointer, while for a dynamically sized space an 
additional four bytes are required per dimension to store the strides for iteration. An additional 
benefit of statically sized spaces 
is that the index computations are performed at compile time, as well as 
the determination of the amount of memory required for the space (shared memory on the GPU, 
or allocation from a pool on the CPU). In practice, these differences are minor compared to the 
computation time of kernels, and the \verb|BlockIterator| over both static and dynamic spaces gives 
similar performance, which is improved compared to the \verb|IndexedIterator| in contexts where memory 
is scarce, such as in shared memory. The interface for \verb|BlockIterator| is a subset of the 
\verb|IndexedIterator| interface, with the \verb|IndexedIterator| providing additional functions which return information about the global iteration space, and traits
to determine the type of an iterator.

\begin{lstlisting}[style=cpp,float=t,caption=Basic graph creation API,label={lst:basic_graph}]
Graph g(ExecutionKind::Cpu);
g.emplace([] { std::cout << "A\n"; })
 .then([] { std::cout << "B\n"; })
 .emplace(
    ExecutionKind::Gpu, /* Specify GPU execution. */
    make_node(
      [] (const char* s) { printf("%s", s); }, "C\n"),
    make_node(
      [] (const char* s) { printf("%s", s); }, "D\n"))
 .then([] { std::cout << "E\n"; });
\end{lstlisting}

%=================================================================
% GRAPHS & SCHEDULING
%=================================================================
\subsection{Graphs}\label{sec:graph}

\begin{figure}[!t]
\centering
\includegraphics{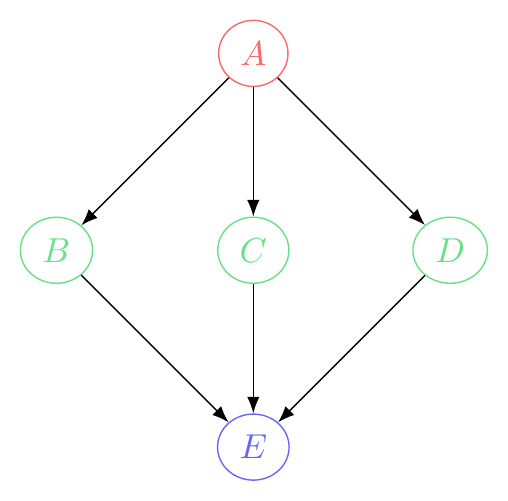}
\caption{Graph structure for the code shown in \rlist{lst:basic_graph}. Colours are used to illustrate
    nodes on the same level. Node A is the root node, on level 0. A, B, and C may execute in parallel, 
    as soon as A completes, and E may execute once B, C, and D are all complete.}
\label{fig:basic_graph}
\end{figure}

Graphs in \frameworkSpace allow a complete workflow 
to be described at a high level, from which the dependencies are determined when 
the graph is built. While the overhead of determining the graph-dependencies is minimal, we incur it at the time of graph creation
because a graph is typically built once but executed many times. The graph is then submitted to an executor, 
which executes the nodes of the graph in parallel such that the dependencies are met. 
Operations are emplaced onto the graph, and can be divided into those on tensor and non-tensor types. 
In this context, a graph is a Directed Acyclic Graph (DAG), where nodes represent the computations (kernels) to be 
performed and edges represent the dependencies between the computations. We define a level in the graph 
to contain all nodes which have a dependency on a node in the previous level, with all nodes 
having no dependencies being the first level. Nodes on the same level may execute in parallel, as soon as their
dependencies are met. The \verb|Graph| 
constructor takes an argument for the default type of execution space,
which can also be overridden for each operation,
as either \verb|ExecutionKind::Cpu| or \verb|ExecutionKind::Gpu|. A simple example
of the levels in the graph is illustrated in \rfig{fig:basic_graph}

\subsubsection{Nodes}

The nodes in a graph are defined using a \verb|Node| which stores the callable defining the 
node operation, as well as any arguments which are required to invoke the callable. To allow callables 
with any signature, type erasure is used to store a generic executor in the node, which is different from most 
existing solutions that require callables to have a specific signature. In addition to the 
callable object, the node stores information to allow it to be accessed and modified after the graph has been created.

\subsubsection{Operations on non-tensor types}

In some task-based models, nodes are added to the graph and then the dependencies 
are specified. However, we take a different approach, where the graph is defined more explicitly and concisely. 
We also provide functionality for modifying dependencies between individual nodes 
after graph creation by allowing nodes to be searched for by the node name. 
The operations which add non-tensor based operations to a graph are \verb|emplace()| and \verb|then()|.

\paragraph{Emplace}

Emplace adds a node onto the graph, with the dependencies of the node being any nodes in the 
previous level. To allow multiple nodes to be placed in the same level (such that they execute in 
parallel), emplace can be called multiple times, or by using the \verb|make_node()| function which allows
multiple callables to be added through a single call to \verb|emplace()|. \rlist{lst:basic_graph} shows 
example usage for the \verb|emplace()| interface.

\paragraph{Then}

The \verb|then()| method creates a dependency in the graph, by indicating that 
the work defined for a new node must be run after previously emplaced nodes. The new node is added as a successor of all nodes in
the previous level, and will only be run after \textit{all} its dependents
have finished. \rlist{lst:basic_graph} shows the code for creating a graph and \rfig{fig:basic_graph} 
shows the structure of the corresponding graph.

\begin{lstlisting}[style=cpp,float=t,caption=Emplace then split example,label={lst:emplace_then_split}]
Tensor<double, 1> x({1}, {4}, size);

Graph g(ExecutionKind::Cpu)
g.emplace([] { std::cout << "A\n"; }),
 .emplace([] { std::cout << "B\n"; })
 .then_split(
   ExecutionKind::Gpu,
   [] ripple_host_device (auto& it) {
     *it = 1.0;
   }, x);
\end{lstlisting}

\subsubsection{Operations on tensor types}

For tensor operations, data may be spread across multiple devices, and therefore a
node which defines its operation on a tensor will, in most cases, have the partitions of the tensor
execute in parallel. To reduce overall latency, the kernel must ideally be submitted asynchronously
to the device on which the partition of the tensor data exists, so that it can begin execution as soon
as all dependencies are met. If a single node were to define the 
operation on the tensor then the submission of the kernel to the multiple devices would be sequential
(since if there is a single node running on a single thread, but multiple kernels are required to be
submitted for each partition, the submissions must be serial). To avoid this, when a kernel is emplaced
onto a graph which operates on a tensor, a node is automatically created for each partition in the tensor, with
each node being on the same level. This allows the kernel associated with each node to be submitted to the device
in parallel, using any number of the available GPU cores in the system. This is more challenging to achieve when the tensor has padding, since memory transfer nodes are required to copy the 
padding data from another device, which may need to complete before the node with the kernel is executed. 
We provide utilities to allow the user to achieve as much parallelism as possible by allowing different requirements for the padding data transfer to be specified, described
in detail in \rsec{sec:data_dependencies}.
For adding nodes which operate on tensors, the \verb|split()| and \verb|then_split()| methods are 
provided, so named to indicate that the operation is split into multiple nodes for each of the 
tensor partitions. 

\paragraph{Split}
\begin{lstlisting}[style=cpp,float=b,caption=Basic split SAXPY example,label={lst:basic_split_saxpy}]
Tensor<double, 1> x({4}, size);
Tensor<double, 1> y({4}, size);

Graph g(ExecutionKind::Gpu);
g.split(
  [] ripple_host_device (
  auto& x_it, auto& y_it, double a) {
    *y_it = a * (*x_it) + (*y_it);
  },
  x, y, 4.0);
\end{lstlisting}
On nodes emplaced for computations on tensor data, \verb|split()| has the same effect 
as \verb|emplace()| in that it will emplace nodes onto the graph at the current level, with each 
of the nodes having a dependency on the nodes in the previous level. If the tensor does 
not require any padding, \verb|split()| creates a node per tensor-partition, and each of 
these nodes may execute concurrently. \rlist{lst:basic_split_saxpy} illustrates code which defines
a graph to perform the well-known saxpy example using four GPUs. 

\begin{figure}[!t]
\centering
\includegraphics{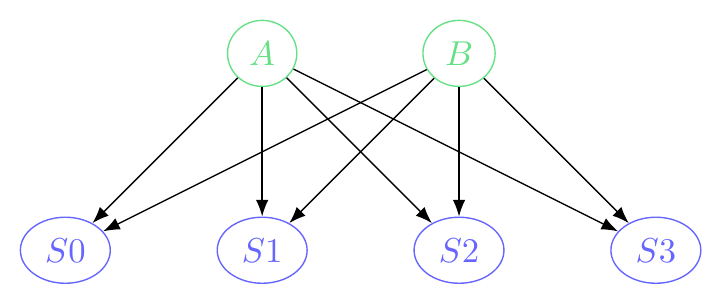}
\caption{Graph structure for the code shown in \rlist{lst:emplace_then_split}. Nodes S0-3 represent the 
    nodes generated for the split operation.}
\label{fig:emplace_then_split}
\end{figure}

\paragraph{Then Split}

The \verb|then_split()| method allows dependencies to be created between operations on previously emplaced 
nodes and the nodes generated by subsequently splitting a tensor. The method specifies that the new operation on each partition of the split tensor must happen after any previous operations in the graph. 
If the previous operation was a \verb|split()| or \verb|then_split()| operation, then dependencies are only 
added between nodes associated with the \textit{same} partitions in the tensor. This reduces the overall 
required connectivity of the resulting graph by reducing the number of synchronization points. 
If a \verb|then_split()| operation is added after a non-tensor operation, then each node generated 
depends on \textit{all} operations in the previous level. This is illustrated in 
\rfig{fig:emplace_then_split} which shows the graph generated by \rlist{lst:emplace_then_split}. 

\paragraph{Reduction} \label{sec:reduction}

\begin{figure}[!t]
\centering
\includegraphics{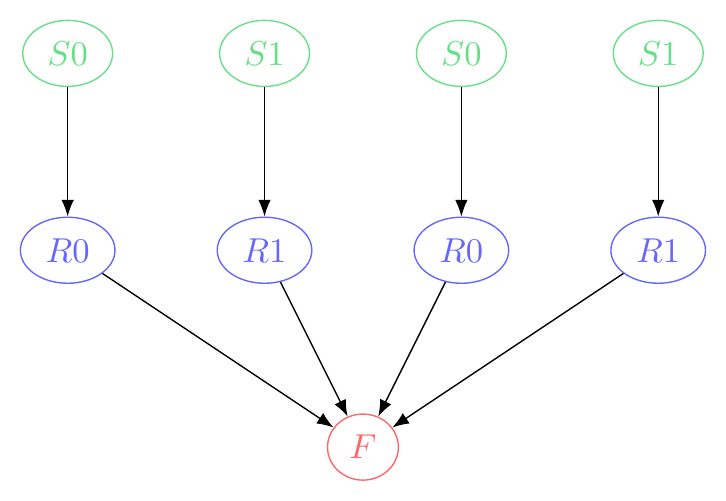}
\caption{Graph structure for the code shown in \rlist{lst:split_then_reduce}. Nodes S0-3 represent the 
    nodes generated for the split operation, R0-3 the reduction operation, and F the operation which 
    sets the reduction as complete. The dependencies allow, for example, for the possibility that
    R0 completes before S1 ends, depending on the dependencies of S0, if this graph were part of a larger
    graph.}
\label{fig:split_then_reduce}
\end{figure}

\begin{lstlisting}[style=cpp,float=b,caption=Split then reduce,label={lst:split_then_reduce}]
Tensor<double, 1> x({4}, size);
auto result = make_reduction_result<double>(0.0);

Graph g(ExecutionKind::Gpu);
g.split(
  [] ripple_host_device (auto& it) {
    *it = 1.0;
  },
  x)
  .then_reduce(x, result, SumReducer());
  
// value = size
auto value = result.value();
\end{lstlisting}

The \verb|reduce()| and \verb|then_reduce()| operations operate specifically on tensors, and reduce the 
data in the tensor into a \verb|ReductionResult<T>| type which stores the result of the reduction as 
well as a flag indicating whether the reduction is complete. The flag is required because for
tensors where data is spread across multiple devices the reduction for each partition of the
tensor may happen asynchronously. 
This design allows each partition to perform the reduction of its data as soon as its dependencies 
are met, so that the entire reduction across multiple devices finishes sooner than if the 
reduction was started once \textit{all} nodes in the previous level were complete. It also 
allows subsequent nodes associated with the partition to continue if they do not have any dependency 
on the result of the reduction. Similarly to \verb|then_split()|, for nodes in the previous 
level that are associated with the \textit{same} partitions on which the reduction will be performed, there 
is only a dependency between the nodes associated with the same partitions, and not between the node 
performing the reduction and all other nodes in the previous level. \rlist{lst:split_then_reduce} shows an 
example for setting each value in a tensor and then performing a reduction sum, while the
corresponding graph is shown in \rfig{fig:split_then_reduce}. The operation for the reduction on the quarter 
of the domain represented by R0 in the graph depends only on the split operation S0 which operates on the 
same data, thus R0 can start as soon as S0 finishes, increasing overall parallelism. Additionally,
the reduction is performed in shared memory on the GPU, which improves performance and only reads the tensor 
data from global memory, so there is no write dependency in subsequent operations, allowing other operations 
on the partitions to be added to the graph at the same level.

\subsubsection{Synchronization}\label{sec:sync}

For larger graphs, some paths may progress more quickly than others, which may then need to be
synchronized. To allow for all pending nodes to complete before any partition
continues, the \verb|sync()| function, with the same signature as
\verb|emplace()| and \verb|split()|, can be used to emplace a node onto the graph which synchronizes all work
from both domains, and then executes the callable for the \verb|sync()| node.

\begin{lstlisting}[style=cpp,float=t,caption=Conditional map reduce with initialization,label={lst:map_reduce}]
Tensor<double, 1> x({4}, size);
auto result = make_reduction_result<double>(0.0);

Graph init(ExecutionKind::Gpu);
Grpah map_reduce(ExecutionKind::Gpu);
init.split(
  [] ripple_host_device (auto& it) {
    *it = 4.0;
  },
  x);
 
map_reduce
  .split([] (auto& it) { *it -= 1.0;})
  .then_reduce(x, result, SumReducer())
  .conditional([&result] {
    return result.value_and_reset() != 0.0;
  });
  
init.then(map_reduce);
\end{lstlisting}

\subsubsection{Subgraphs}\label{sec:subgraphs}

To enable better separation of code, and overall flexibility in the graph creation API, we allow 
graphs to be added using the \verb|emplace()| and \verb|then()| operations, which adds the graph passed to the 
operation as a subgraph in the current graph, which will either execute in parallel with the nodes in the last 
level (for \verb|emplace()|) or after the nodes in the last level (for \verb|then()|). For complex 
graphs, this allows each of the subgraphs to be implemented separately, and then to be combined in a final step 
which defines the connectivity between the subgraphs.

To ensure that the resulting graphs are correct, the nodes in the last level of the graph onto which the subgraph 
is emplaced need to be connected to the nodes in the first level of the subgraph. For nodes on non-tensor types 
the process is simple since each node in the last level of the graph is connected to each node in the first level 
of the subgraph. However, as mentioned in \rsec{sec:reduction}, when the nodes in the levels operate on the same 
tensor partition then the node in the subgraph should only be connected to the node which corresponds to the same 
partition in the tensor, as well as any other non-tensor nodes in the previous level. If this is not the case, the operations 
on the tensor partition will wait longer than necessary due to an unnecessary dependency between the 
node and another tensor node which operates on a different partition.

\subsubsection{Conditional execution}\label{sec:conditional}

\begin{figure}[!t]
\centering
\includegraphics[width=.85\linewidth]{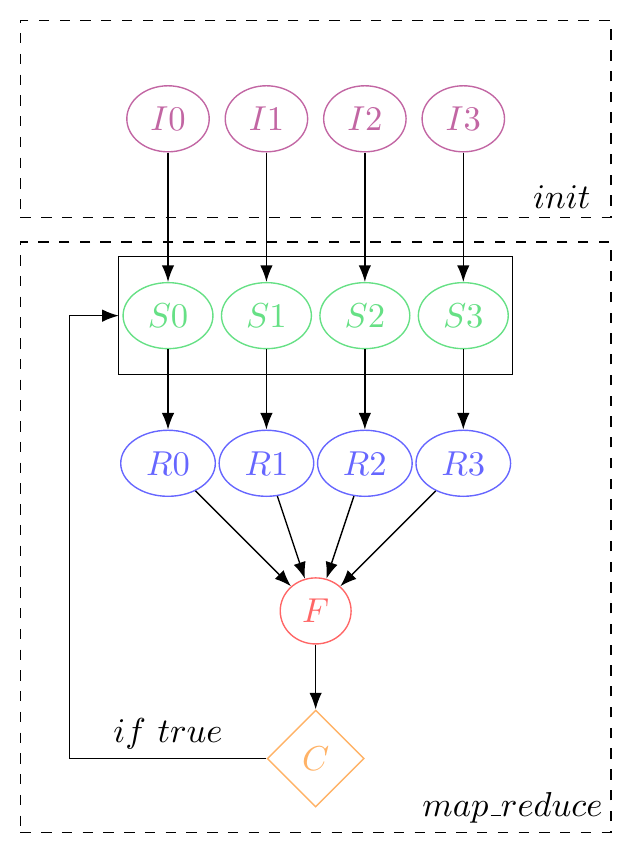}
\caption{Graph structure for the code shown in \rlist{lst:map_reduce}, indicating both graphs from the code listing, and the flow of control. }
\label{fig:map_reduce}
\end{figure}

There is a trade-off between static and dynamic specification of a graph; if a graph is specified 
statically then compile-time optimizations can reduce its complexity so that the execution
is more efficient, however, for most real-world applications, graphs likely need some form of dynamic 
behaviour at runtime. We provide dynamic behaviour through conditional operations 
that can be emplaced onto a graph, which determine whether or not the graph should be executed.
A conditional node takes a predicate which returns \verb|true| 
if the graph should execute, and \verb|false| otherwise. Along with the subgraph functionality 
described in \rsec{sec:subgraphs}, this allows graphs of arbitrary complexity to be created which have completely dynamic 
runtime behaviour. The expressiveness of this functionality can be seen by combining the \verb|split()| and 
\verb|then_reduce()| operations on tensor data to provide a simple implementation of the MapReduce \cite{Dean2008} framework. 
An example is shown in \rlist{lst:map_reduce} and the corresponding graph in \rfig{fig:map_reduce}.

%=================================================================
% EFFECTS OF DATA LAYOUT ON PERFORMANCE
%=================================================================
\subsection{Tensor Data Dependencies}\label{sec:data_dependencies}

As described in \rsec{sec:graph}, when nodes are added to the graph for tensor operations it is important to reduce 
the connectivity of the resulting graph. This allows operations on independent partitions of the tensor to begin execution 
as soon as possible, reducing synchronization during execution of the graph. The complexity of the graph, 
however, depends on the how the kernel modifies the partition data associated with a node---whether the 
operation on the data requires concurrent or exclusive access and 
whether the tensor has padding accessed by the kernel.

The simplest case is a tensor operation requiring no padding, since it cannot require data from
neighbouring partitions. This does not cause any ordering dependency between the node and 
any subsequent nodes which operate on the same data. Thus, when a tensor is passed to one of the methods for emplacing 
nodes onto the graph we assume this is how the tensor is accessed. For any other type of access, one of the methods for 
specifying the access requirement when the operation uses padding must be used, which ensures that the correct dependencies
are added to the graph to avoid race conditions.

\subsubsection{Concurrent access with padding}

\begin{figure}[!t]
\centering
\includegraphics[width=.96\linewidth]{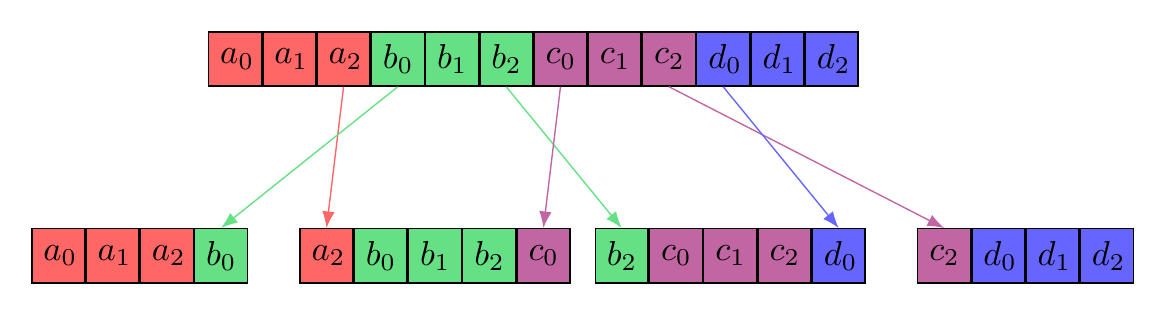}
\caption{One-dimensional tensor data. Colours for data represent that data resides on the same device, colours for edges 
    represent the device which performs the operation to move data along the edge.}
\label{fig:concurrent_access}
\end{figure}

\begin{lstlisting}[style=cpp,float=b,caption=Double buffered concurrent read example,label={lst:concurrent_padding}]
Tensor<double, 1> x({4}, padding, size);
Tensor<double, 1> y({4}, size);

auto callable = 
  [] ripple_host_device (auto&& in, auto&& out) {
    *out = *in.offset(ripple::dim_x, 1) 
         - *in.offset(ripple::dim_x, -1);
  };

graph.split(
  callable, ripple::concurrent_padded_access(x), y);
\end{lstlisting}

\begin{figure}[!b]
\centering
\includegraphics[width=.96\linewidth]{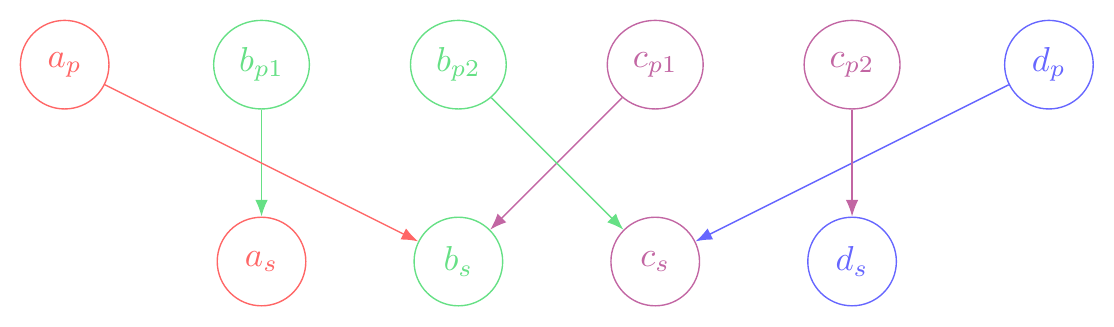}
\caption{Graph for transferring concurrently accessed data with padding. Edge colours correspond to the data movement 
    in \rfig{fig:concurrent_access}, where subscripts $p$ and $s$ correspond to the padding and split operations, 
    respectively. Nodes of the same colour in different levels can execute concurrently.}
\label{fig:concurrent_access_graph}
\end{figure}

\begin{figure}[!t]
\centering
\includegraphics[width=.96\linewidth]{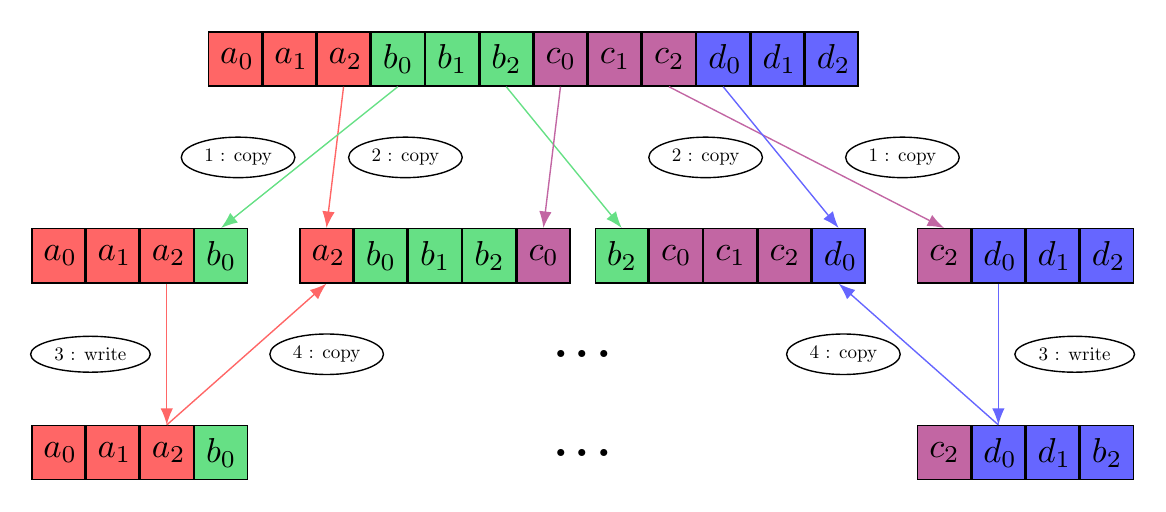}
\caption{One-dimensional tensor data which shows a sequence where a potential race condition occurs between
    operations 2 and 3, which may result in the padding data $a_2$ being the updated value. 
    Exclusive padded access, results in a graph which ensures that operation 2 \textit{happens before}
    operation 3, removing the race condition so that the transferred $a_2$ is the original value.}
\label{fig:tensor_write}
\end{figure}

\begin{lstlisting}[style=cpp,float=b,caption=Single data exclusive write example,label={lst:exclusive_padding}]
Tensor<double, 1> x({4}, padding, size);

auto callable = 
  [] ripple_host_device (auto&& in_out) {
    *in_out = *in_out.offset(ripple::dim_x, 1) 
            - *in_out.offset(ripple::dim_x, -1);
  };

graph.split(
  callable, ripple::exclusive_padded_access(x));
\end{lstlisting}

\begin{figure}[!b]
\centering
\includegraphics[width=.96\linewidth]{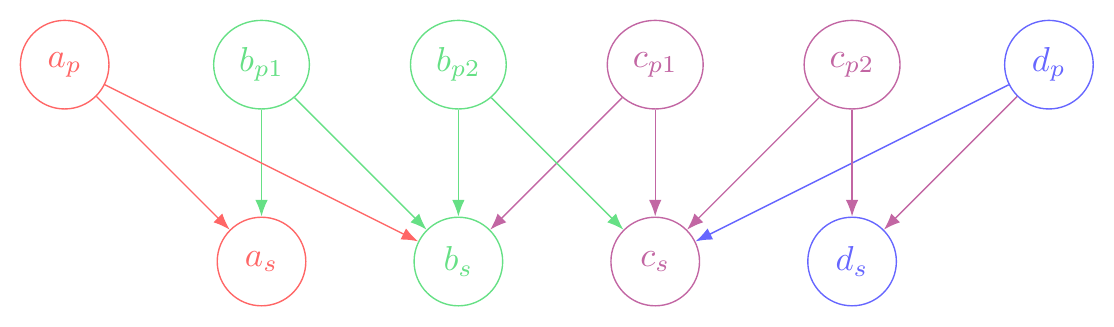}
\caption{Graph for a one dimensional tensor which requires padding and where the subsequent operations 
    write to the data, where subscripts $p$ and $s$ correspond to the padding and split operations, 
    respectively. Nodes of the same colour in different levels cannot execute concurrently.}
\label{fig:tensor_write_graph}
\end{figure}

\begin{figure*}
    \centering
    \includegraphics[width=0.99\linewidth]{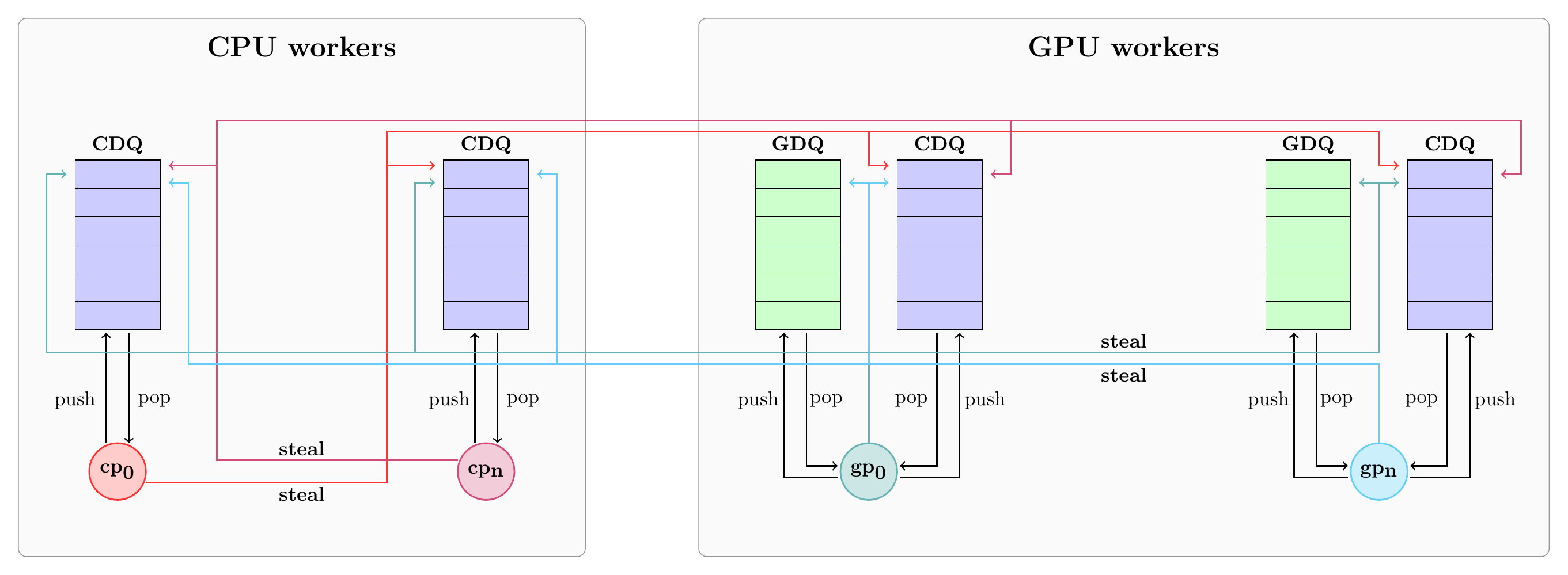}
    \caption{Design of our heterogeneous scheduler. coloured paths illustrate the possibility for a worker to steal from the
        queue to which the path leads.}
    \label{fig:scheduler}
\end{figure*}

When the tensor partition data can be accessed concurrently (i.e~there is no write to data which is used by a kernel on another partition) 
by a kernel and the access \textit{will} use the padding data, then the \verb|concurrent_padded_access()| modifier needs
to be used when the kernel is emplaced onto the graph. 
Consider the one-dimensional tensor data in \rfig{fig:concurrent_access}, where the data is split across four devices and that the operation to be performed is the difference computation in \rlist{lst:concurrent_padding}. 
The figure illustrates the padding which needs to be copied from neighbouring devices before the difference 
computation---which accesses the padding---can begin. Using the \verb|concurrent_padded_access()| modifier, as shown in 
\rlist{lst:concurrent_padding}, results in the associated graph shown in \rfig{fig:concurrent_access_graph}. An additional node is inserted for padding data transfer operation from a partition to a neighbour partition, which ensures 
that the subsequent nodes can run concurrently as soon as their padding operations are complete. 
The padding operations \textit{from} a node (\rfig{fig:concurrent_access_graph} ($x_p$)) can happen on 
different streams than the kernels for \textit{the same} node ($x_s$) and, if the GPU supports it, can overlap in time. For example, in
\rfig{fig:concurrent_access_graph}, the operation $a_p$ could happen in parallel with $a_s$, since the data copied by $a_p$ is
not modified in $a_s$. For the one-dimensional case, the highest padding cost is two copies (nodes $b_{p1}, b_{p2}, c_{p1}$ and 
$c_{p2}$ in \rfig{fig:concurrent_access_graph}), however, for higher dimensional data the number of copies is significantly more
if the data is split across all dimensions (8 for two dimensions, and 26 for three dimensions). While splitting data across 
fewer dimensions results in a lower number of data transfers, the best partitioning strategy is problem-dependent, hence why we make it easy for the user to specify.

\subsubsection{Exclusive access with padding}

When the tensor partition data needs to be accessed exclusively (i.e.~there is a write to data which is required by a kernel on another partition) 
by the kernel for the node, and the kernel \textit{will} use padding data, then the
\verb|exclusive_padded_access()| modifier needs to be used when the operation is emplaced onto the
graph, as the resulting graph requires additional connections to prevent a race condition between data write-access in the kernel and the memory transfer operation of the same data to a neighbour which requires reading it for padding. This
is illustrated in \rfig{fig:tensor_write} for the graph in \rfig{fig:concurrent_access_graph}. Since all the operations to copy the padding data for the leftmost partition can execute in parallel, it is possible that the copy 
(operation 1) for its padding completes before the copy (operation 2) for the second leftmost 
partition's padding begins. Additionally, since for the graph in \rfig{fig:concurrent_access_graph} the split kernel
and the data transfers for the same partition can happen concurrently, if dependencies for a write operation 
are not added in \rfig{fig:concurrent_access_graph}, then there will be a race between operations 2 and 3 in 
\rfig{fig:tensor_write}. 
This race happens for all the padding data, as illustrated by the rightmost data in \rfig{fig:tensor_write}.
To avoid this, the \verb|exclusive_padded_access()| modifier specifies that a kernel will use padding data and requires
exclusive access, allowing the library to include the required dependencies in the graph, as shown in \rfig{fig:tensor_write_graph}. The split operations on the second level, in addition to the dependencies of the
padding operations from their neighbours, have dependencies on the padding operations to their neighbours. Such access
is usually required when there is only a single instance of the data, which is updated within a kernel, as in \rlist{lst:exclusive_padding}.

\begin{algorithm}[b!]
\caption{Worker thread main loop}
\begin{algorithmic}[1]\label{alg:worker_main_loop}
\renewcommand{\algorithmicrequire}{\textbf{Input:}}
\REQUIRE s : worker state
\REQUIRE t : worker thread
  \WHILE {(!s.must\_shutdown())}
    \IF {(s.paused())}
      \STATE s.suspend()
      \STATE \textbf{continue}
    \ENDIF
    \IF {(execute\_work(s, s.main\_priority()))}
      \STATE \textbf{continue}
    \ENDIF
    \STATE execute\_work(s, s.secondary\_priority())
  \ENDWHILE
\end{algorithmic} 
\end{algorithm}

\subsubsection{Shared access}

We also provide access specifiers which create and copy the tensor data into shared memory, which can improve performance for
some kernels. When the data has no padding, the \verb|in_shared()| specifier can be used, while for data which additionally requires
padding, the \verb|exclusive_padded_acccess_in_shared()| and \verb|concurrent_padded_access_in_shared()|
modifiers can be used to create these dependencies.

%=================================================================
% Scheduling
%=================================================================
\section{Scheduling}\label{sec:scheduling}

\begin{algorithm}[b!]
\caption{execute\_work(s, q)}
\begin{algorithmic}[1]\label{alg:execute_node}
\renewcommand{\algorithmicrequire}{\textbf{Input:}}
\REQUIRE s : worker state
\REQUIRE q : queue type
\renewcommand{\algorithmicrequire}{\textbf{Output:}}
\REQUIRE bool : if execution succeeded
  \STATE node $\gets$ s.pop(q)
  \IF {(node)}
    \IF {(node.try\_execute())}
      \STATE s.increment\_processed\_nodes()
      \RETURN true
    \ENDIF 
    \STATE s.push(node, q)
  \ENDIF
  \RETURN steal(s, q)
\end{algorithmic} 
\end{algorithm}

For scheduling, we use a modified form of the well-known work stealing algorithm. We provide an overview of the
algorithm, and our extensions of the algorithm to multiple domains of execution---CPUs and GPUs.

\subsection{Work Stealing}\label{sec:work_stealing}

Work stealing is a common pattern for task-based parallelism and has been used successfully in
\cite{Leijan2009, Lee2010, huang2020}. A group of worker threads is spawned, each with its own
queue of work from which tasks can be executed. When a thread's queue becomes empty, it \textit{steals} work from the queues
of other worker threads \cite{Blumofe1999, Arora1998}. Once a worker thread completes execution of a task it looks to
the successor of the completed task as the next candidate for execution, which can be executed if all the dependencies 
of the task are met \cite{Agrawal2010}. The pool of worker threads continues to execute until all queues are empty, or 
the program terminates. 

Most existing research on work stealing has focused on a single domain, which has primarily been the CPU in the context of large 
task-based systems which utilise hundreds to thousands of CPU cores. For heterogeneous domains, however, there is little
work on using work stealing to achieve good parallel efficiency across domains. In such a context, worker threads could submit tasks for 
execution on either the CPU or the GPU, and the scheduling algorithm needs to ensure that the dependencies in the graph 
are met for the execution of both domains, while still being efficient. The most recent work for heterogeneous 
domains is that of \cite{huang2020}. However, their work focuses on the execution of large graphs on the CPU. While
it does allow nodes to contain GPU work, it does not have knowledge of how data is accessed by kernels, and therefore cannot
allow for determining data dependencies for GPU-based workflows, instead requiring
that the user write all GPU code within the graph nodes. We follow their heterogeneous scheduler approach, 
but modify it to better support our higher-level GPU abstractions to allow good performance for GPU-based workflows.

\subsection{Design}\label{sec:sched_design}

\begin{algorithm}[t!]
\caption{try\_execute()}
\begin{algorithmic}[1]\label{alg:try_execute}
\renewcommand{\algorithmicrequire}{\textbf{Require:}}
\REQUIRE n : node to execute
\renewcommand{\algorithmicrequire}{\textbf{Output:}}
\REQUIRE bool : if execution succeeded
  \IF {(n.dependents.load() $\neq$ 0)}
     \RETURN false
  \ENDIF
  \STATE n.dependents $\gets$ n.inital\_dependents
  \STATE n.execute()
  \FOR{\textbf{each} s in n.successors}
    \STATE s.dependents.fetch\_sub(1)
  \ENDFOR
  \RETURN true
\end{algorithmic} 
\end{algorithm}

We use the general work stealing idea, with a pool of worker threads, however, we additionally give each worker
thread a domain priority for the type of work it primarily schedules---as either CPU or GPU. The worker
threads in the algorithm are adaptive, sleeping when there is not enough work for all threads in the domain, 
and waking when the workload increases, which is determined when nodes are scheduled for execution. 
Each worker thread stores state information to facilitate its suspension: 
the domain priority, the number of processed nodes, and the queue of nodes. The main loop for the worker 
threads in our scheduler is shown in \ralg{alg:worker_main_loop}.

By default, we set the number of threads with GPU priority (GPUP) to be one more than the number of physical GPUs in 
the system, and the number of threads with CPU priority (CPUP) to be the number of remaining physical CPU cores in 
the system. In all benchmarks, using both more or less GPUP threads than this reduced performance due to 
either requiring more CUDA context switching and stream synchronization than necessary, when more threads were used, 
or too little parallelism, when fewer were used.

\begin{algorithm}[t!]
\caption{steal(s, q)}
\begin{algorithmic}[1]\label{alg:steal}
\renewcommand{\algorithmicrequire}{\textbf{Input:}}
\REQUIRE s : worker state
\REQUIRE q : queue type
\renewcommand{\algorithmicrequire}{\textbf{Output:}}
\REQUIRE bool : if execution succeeded
  \STATE steal\_id $\gets$ s.thread\_id
  \FOR {(i $\in$ [0, s.steal\_attempts])}
    \STATE steal\_id $\gets$ s.get\_steal\_id(s.steal\_strategy)
    \IF {(steal\_id == s.thread\_id)}
      \STATE \textbf{continue}
    \ENDIF
    
    \STATE node $\gets$ thread\_states[steal\_id].steal(q)
    \IF {(node)}
      \IF {(node.try\_execute())}
        \STATE s.increment\_processed\_nodes()
        \RETURN true
      \ENDIF 
      \STATE s.push(node, q)
    \ENDIF
  \ENDFOR
  \RETURN false
\end{algorithmic} 
\end{algorithm}

We must ensure that when multiple workers try to 
steal the same task from the same queue, it is only possible for \textit{a single worker} to gain access to the task. Thus the main data structure for the scheduler is a lock-free 
implementation of a work stealing deque \cite{Chase2005, Le2013}, where the thread to which the queue belongs can 
push and pop from one end, while other threads can steal from the other end. GPUP threads have queues for both CPU 
(CDQ) and GPU (GDQ) work, and can push and pop onto either of their own queues (prioritising GDQs) or steal from 
the GDQ or CDQs (again prioritising the GDQs) of any other threads. CPUP threads, however, have only a single work
stealing deque (CDQ) and can only push and pop from it, but can steal from any other CDQ, regardless of the domain 
priority of the thread to which the queue belongs. This allows all CPU cores to perform multi-threaded CPU 
work, but only a subset to perform submission and synchronization of GPU work, allowing an increase in the
level of CPU parallelism when some GPU kernels are significantly longer executing than other CPU kernels.

In the work stealing deque for a worker thread, we encapsulate the work to be executed by the executor in a \verb|Node|
class---which represents a node in a graph to be executed---and store the nodes in the deque. The node contains information 
about its connectivity in the graph, via an atomic counter which stores the number of dependents, and a pointer to its 
successor, whose dependents counter needs to be decremented when the node completes its work, allowing the graph 
dependencies to be maintained. Within the \verb|execute_work| function of \ralg{alg:worker_main_loop}, a worker pops 
a node off one of its queues, tries to execute it, and then either exits, upon success, or tries to steal from another 
queue. This is shown in \ralg{alg:execute_node}, while the algorithm for attempting to execute a node's work is shown 
in \ralg{alg:try_execute}. On line 5 of \ralg{alg:try_execute} we restore the initial dependency count of the node 
before it finishes executing, which facilitates the conditional execution within graphs as described in 
\rsec{sec:conditional}.

When a node fails to execute due to unmet dependencies, it is pushed back onto the queue of the worker which
tried to execute it, which is cheap
due to there being no contention with other threads for pushing and popping onto the one side of the work stealing deque. This
allows other nodes to execute the node work if the dependencies are met while the node is stealing from other threads. When a node
attempts to steal from other nodes, there are a number of strategies which can be used, and any number of attempts can be made before
giving up and going back to a node's own queues. Three stealing strategies are: the random strategy which randomly chooses a queue from another thread, the round robin strategy which chooses a queue by increasing
index starting from one more than the stealing thread's index, wrapping around, the topological strategy which chooses a queue based 
on the cache sharing properties of the thread, trying to choose queues from threads which share caches. Currently, we employ both 
the random and round robin strategies, however, we have seen that the performance of round robin is far superior to random stealing, 
likely due to the lower likelihood of two threads trying to steal from the same queue, leading to less contention. The stealing algorithm
is shown in \ralg{alg:steal}.

For a node whose work needs to be executed on the GPU, it is important that the resulting graph is efficient, i.e.~not introducing
points where the GPU is unnecessarily idle because of over synchronization. Thus, when calling the CUDA API we use the asynchronous version of all memory related functions to ensure that 
threads are not blocked, and rely on atomic operations in the work stealing deque and node execution function to ensuring
scheduling dependencies. Lastly, we introduce stream synchronization when a graph node requires that GPU work completes before 
executing the node. 
We use multiple streams per GPU for copy and compute to allow memory transfer and kernel execution to happen asynchronously where 
it is possible (for example \verb|concurrent_padded_access()|), but place operations onto the same compute stream when 
it is not, since operations on the same stream are performed synchronously in the order in which they are submitted. This produces compact GPU pipelines, resulting in low overall overhead and good scaling performance, as illustrated in the results section.

\begin{table}[t!]
    \resizebox{\columnwidth}{!}{%
    \centering
    \begin{tabular}{l|c|c|c}
        \hline
        Name                & AWS p3.2xlarge                 & AWS p3.4xlarge                             & AWS p3.8xlarge    \\\hline
        OS                  & \multicolumn{1}{c}{}      & \multicolumn{1}{c}{Centos 7}          &                       \\
        CPU                 & \multicolumn{1}{c}{}      & \multicolumn{1}{c}{Intel Xeon E5-2686 v4}     &                       \\
        CPU Frequency       & \multicolumn{1}{c}{}      & \multicolumn{1}{c}{2.3 GHz}           &                       \\
        GPU                 & \multicolumn{1}{c}{}      & \multicolumn{1}{c}{NVIDIA V100-SXM2}       &                       \\
        GPU Architecture    & \multicolumn{1}{c}{}      & \multicolumn{1}{c}{Volta (sm 7.0)}    &                       \\
        CUDA Cores          & \multicolumn{1}{c}{}      & \multicolumn{1}{c}{5120}              &                       \\
        GPU Frequency        & \multicolumn{1}{c}{}      & \multicolumn{1}{c}{1.53 GHz}               &                       \\
        GPU Memory          & \multicolumn{1}{c}{}      & \multicolumn{1}{c}{16 GB per GPU}     &                       \\\cline{2-4}
        CPU Cores           & 8                         & 32                                    & 64                    \\
        Memory              & 61 GB                     & 244 GB                                & 488 GB                \\
        GPUs                & 1                         & 4                                     & 8                     \\\hline
    \end{tabular}}
    \caption{System configuration used to generate results.}
    \label{tab:configuration}
\end{table}

%=================================================================
% EFFECTS OF DATA LAYOUT ON PERFORMANCE
%=================================================================
\section{Fine-grained Performance Evaluation}\label{sec:data_layout_results}

The performance tests in 
this section are all performed on a single GPU to allow a comparison against implementations
of similar kernels using other frameworks or libraries designed for 
fine-grained parallelism, such as Kokkos \cite{EDWARDS2014}.
The hardware configurations of the AWS compute nodes are shown in 
\rtab{tab:configuration}, where we use the p3.2xlarge node for single GPU tests, and all larger node 
types for the scaling tests in \rsec{sec:scaling_results}.

When implementing the kernels in Kokkos we follow 
the examples presented in \cite{EDWARDS2014}, as well as the examples in the Kokkos repository. The implementations of the kernel function objects and structs for Kokkos and 
\frameworkSpace are very similar, with the exception that for \frameworkSpace there is a
template parameter for the layout kind for any struct which is used in a benchmark.

\subsection{SAXPY}\label{sec:saxpy}

The first benchmark is SAXPY which computes
\begin{equation*}
    \textbf{y} = a * \textbf{x} + \textbf{y}
\end{equation*}
Its minimal computational requirements make
it a good example for determining the overhead of the iterator functionality provided by \framework. 
We compare the performance of \frameworkSpace against both cuBLAS~\cite{Cuda2020} and Kokkos, with the timings shown in \rtab{tab:saxpy}. For 1 billion vector elements, \frameworkSpace takes 56\% of the time taken by Kokkos, and 99.2\% of that taken by cuBLAS.

\begin{table}[t!]
    \centering
    \resizebox{0.85\columnwidth}{!}{%
    \begin{tabular}{l|cccc}
        \hline
        Size    & cuBLAS     & Kokkos     & Ripple     & Ripple NBC                 \\\hline
        1M      & 0.0749     & 0.0421     & 0.0327     & 0.0291                     \\
        10M     & 0.2316     & 0.2722     & 0.1704     & 0.1119                     \\
        100M    & 1.5286     & 2.5773     & 1.5242     & 1.4622                     \\
        200M    & 2.9785     & 5.1334     & 3.0356     & 2.9088                     \\
        500M    & 7.3153     & 12.826     & 7.5261     & 7.2347                     \\
        1B      & 14.556     & 25.618     & 15.048     & 14.439                     \\\hline
    \end{tabular}}
    \caption{Results of the SAXPY computation. The \frameworkSpace NBC column is the performance
        using \frameworkSpace without checking the boundary condition. All times are in milliseconds.}
    \label{tab:saxpy}
\end{table}

Profiling indicates that \frameworkSpace has a small amount of overhead for the iterator abstractions, due to the checking of the validity of the iterator. The overhead
is in the range of 3-4\%, but this is for a simple kernel which loads data into
registers, performs a single fused multiply-add (FMA), and loads the data back to global memory. This kernel uses a one dimensional iterator, so the boundary check is only performed once. For higher dimensional data spaces the overhead would be higher, since the validity of the iterator needs
to be checked for each dimension. For workloads which are more computationally intensive, however, the overhead
would be a smaller relative percentage.

\subsection{Particle Update in Three Dimensions}\label{sec:particle_update}

This kernel computes
\begin{equation*}
    \textbf{x} = \textbf{x} + \textbf{v}\Delta t
\end{equation*}
where $\textbf{x}$ and $\textbf{v}$ are the three-dimensional position and velocity for each particle.
This benchmark is similar to the SAXPY benchmark, but requires significantly more data, which better tests the memory access performance
of \framework. Additionally, the position and velocity data require vector data types with multiple components, which can either be stored 
contiguously or strided. 

\begin{table}[t!]
    \centering
    \resizebox{\columnwidth}{!}{%
    \begin{tabular}{l|ccccc}
        \hline
        Size        & Kokkos    & Ripple (s)    & Ripple (c)    & Ripple (ss)   & Ripple (cs)           \\\hline
        100k        & 0.0255    & 0.0191        & 0.0259        & 0.0221        & 0.0379                \\
        1M          & 0.1444    & 0.1036        & 0.1620        & 0.1343        & 0.2381                \\
        10M         & 1.2743    & 0.8679        & 1.4472        & 1.1834        & 2.0507                \\
        20M         & 2.5612    & 1.7146        & 2.8487        & 2.3441        & 4.0743                \\
        40M         & 5.0522    & 3.4062        & 5.7184        & 4.6741        & 8.1453                \\
        80M         & 10.112    & 6.7914        & 11.452        & 9.3259        & 16.3081               \\\hline
    \end{tabular}}
    \caption{Results of the particle update kernel for a single update for the various layouts in \frameworkSpace and for Kokkos. 
        All times are in milliseconds. s = strided layout, c = contiguous layout, ss = strided layout in shared memory, cs = contiguous layout in shared memory.}
    \label{tab:particle_update}
\end{table}

For a three-dimensional
update, the position and velocity require 12 or 24 bytes each, for single and double precision, respectively. We define a \verb|Particle| class 
similar to the \verb|State| class shown in \rlist{lst:layout_example}, to hold the particle position and velocity and use an 
\verb|update(dt)| method to update the position given $\Delta t$. The results are shown in \rtab{tab:particle_update}. This benchmark is memory 
bound, since for both the position and the velocity, each thread needs to read the six elements, perform an FMA to update the position for each
dimension, and then write the results back to global memory. For contiguous data, coalesced access is not possible, so more 
than a single memory transaction is required per element in each vector. However, when the data is strided, only a single memory transaction is
required per element so the memory access is coalesced for each element, which for this benchmark results in a 33\% improvement over the
contiguously laid out implementation in Kokkos, and a 40\% improvement over the contiguous implementation in \framework.

We also see that using shared memory is slower for this benchmark, due to the increase in
the instruction count of the kernel. For compute architectures prior to Ampere (such as the Volta architecture used
here), loading data into shared memory goes through the register file, both increasing register usage, which may reduce occupancy, as well as
increasing the overall instruction count of the kernel. For a kernel with such a simple computation it
is not necessary to use shared memory, since accessing data from global memory already puts the data into a register, so
a load to and from shared memory reduces the performance. For the recent Ampere architecture, however, there is new functionality for copying data from global
memory to shared memory asynchronously which bypasses the load into a register. This could improve performance of the shared memory results
presented here. There is also more fine-grained control of the synchronization for this type of asynchronous data loading, which allows the compute
cores to be utilised while waiting for the load to complete. We intend to investigate how to take advantage of this for our iterator abstractions.

\begin{table}[t!]
    \centering
    \resizebox{0.8\columnwidth}{!}{%
    \begin{tabular}{l|cccc}
        \hline
        Size$^2$        & Kokkos     & Ripple (s)   & Ripple (c)    & Ripple (ss)       \\\hline
        1k              & 0.2476     & 0.0757       & 0.1065         & 0.0921           \\
        2k              & 0.8259     & 0.2438       & 0.3614         & 0.2979           \\
        4k              & 3.2481     & 0.9101       & 1.3971         & 1.1425           \\
        8k              & 12.919     & 3.5533       & 5.5284         & 4.4941           \\
        16k             & 51.857     & 14.282       & 22.081         & 18.090           \\
        32k             & 215.56     & 59.028       & 86.756         & 74.797           \\\hline
    \end{tabular}}
    \caption{Results of a single computation of the flux difference in two dimensions for the various layouts in \frameworkSpace and for Kokkos. 
        All times are in milliseconds. s = strided layout, c = contiguous layout, ss = strided layout in shared memory, cs = contiguous layout in shared memory.}
    \label{tab:flux}
\end{table}

\subsection{Flux Difference in Two Dimensions}\label{sec:flux_benchmark}

This benchmark computes the flux difference across all faces in two dimensions. It is a typical stencil-type computation which
is used in finite volume and finite element methods. It has a larger computational component and also 
uses data from neighbouring cells in both dimensions. The memory access pattern is therefore more demanding than the previous 
benchmarks, and the kernel requires more registers for the computation. The kernel for the benchmark computes
\begin{equation*}
    \textbf{F} = \sum_i^D F_{i + \frac{1}{2}} - F_{i - \frac{1}{2}}
\end{equation*}
where $D$ is the number of dimensions, $F_{i \pm \frac{1}{2}}$ is the flux at each face of the cell in the dimension $i$, which can 
be computed with any number of methods. We choose to use the FORCE method of Toro~\cite{Toro1996}. Each cell stores 
a \verb$State$ data type for the solution of the Euler equations, with components for density, 
energy, and the velocity in each dimension. The complexity of the computation comes from the significant number of multiplications 
of each of the components in the \verb|State|, which are essentially vector multiplies. The computation of the flux for the state
also needs to be computed in the FORCE method, which requires evaluating the pressure from the other state variables, resulting in
many reads and writes to the state data variables. As this is a two-dimensional test, the FORCE evaluation 
must be performed for all four faces of each cell. Lastly, padding data must be added to the domain so that the FORCE method, which
requires the data from neighbouring cells in the dimension being solved, computes valid results for the first and last cell in both
dimensions. Despite the additional computational complexity, the effects of data layout are
the same as in the previous benchmarks, with strided data giving significantly better performance. The results are shown in \rtab{tab:flux}.
We see that on a large problem \frameworkSpace with the strided layout is 3.65 times faster than Kokkos.

\subsection{Eikonal Equation Solution}\label{sec:fim_benchmark}

\begin{table}[t!]
    \centering
    \resizebox{\columnwidth}{!}{%
    \begin{tabular}{l|ccccc}
        \hline
        Size$^2$    & Kokkos    & Ripple (s)    & Ripple (c)    & Ripple (ss)       & Ripple (cs)           \\\hline
                    & float,    & 5 cells       &               &                   &                       \\\cline{2-6}
        1k          & 0.0991    & 0.0712        & 0.1061        & 0.0592            & 0.0689                \\
        2k          & 0.3375    & 0.2294        & 0.3776        & 0.1835            & 0.2161                \\
        4k          & 1.2554    & 0.8720        & 1.4409        & 0.6788            & 0.8041                \\
        8k          & 5.1286    & 3.4141        & 5.7242        & 2.6383            & 3.1509                \\
        16k         & 20.501    & 13.537        & 22.735        & 10.524            & 12.550                \\\cline{2-6}
                    & double,   & 10 cells      &               &                   &                       \\\cline{2-6}
        1k          & 0.3150    & 0.1509        & 0.3235        & 0.1239            & 0.1756                \\
        2k          & 1.1978    & 0.5374        & 1.2181        & 0.4300            & 0.6341                \\
        4k          & 4.7241    & 2.0738        & 4.7533        & 1.6370            & 2.4578                \\
        8k          & 18.9335   & 8.1661        & 18.590        & 6.4601            & 9.7172                \\
        16k         & 76.112    & 32.405        & 73.924        & 25.632            & 38.812                \\\hline  
    \end{tabular}}
    \caption{Results of the FIM solve for \frameworkSpace and for Kokkos. 
        All times are in milliseconds. s = strided layout, c = contiguous layout, ss = strided layout in shared memory, cs = contiguous layout in shared memory.}
    \label{tab:fim}
\end{table}

%=================================================================
% SCALING
%=================================================================

The previous benchmarks, while having increasing levels of computational complexity, were all memory bound. This benchmark is computationally bound. The Eikonal equation is a special case of the nonlinear Hamilton–Jacobi partial 
differential equations (PDEs), defined on a Cartesian grid with a scalar speed function as
\begin{equation*}
    H(\textbf{x}, \nabla \phi) = | \nabla \phi (\textbf{x}) |^2 - \frac{1}{f(\textbf{x})^2} = 0
\end{equation*}
where $\phi(\textbf{x})$ is a scalar field defined at each cell in the domain with location $\textbf{x}$, and $f(\textbf{x})$ the 
speed at the cell $\textbf{x}$. Uses of the solution of the Eikonal equation include reinitializing a level-set function 
and performing multi-dimensional extrapolation of data. For this benchmark
we set $f(\textbf{x}) = 1$, which is the case for reinitializing a signed-distance function.

The solution of the equation propagates information from source nodes throughout the domain---for example, for levelset reinitialization
the source nodes would be the locations at which the levelset value is zero. Again, there are a number of methods which can
be used to solve the Eikonal equation, such as The Fast Marching Method \cite{Sethian1996}, the Fast Sweeping Method \cite{Zhao2005}, \cite{Zhao2006} and the Fast Iterative Method \cite{Jeong2008} 
have been used for parallel implementations. We choose to use the Fast Iterative Method for this benchmark, and set a
single source node at the centre of the domain. Varying the number of cells in the reinitialization band varies the computational 
expense of the problem, and we therefore perform the computation with different numbers of cells in the band around the source node,
as well as for different levels of precision. The results are shown in \rtab{tab:fim}. With strided layout in shared memory, \frameworkSpace is 1.95 times faster than Kokkos for 5 cells in single precision, and 2.97 times faster for 10 cells in double precision.

Using shared memory gives up to 30\% improvement in performance compared to global when using strided data, while compared to
contiguous data the strided shared memory improvement is closer to 90\%. This difference comes from the fact that the computational
component of the results contributes essentially the same time when using shared memory, regardless of data layout, however, the
transfer between global and shared memory is significantly faster for strided data, as shown previously. The combination of 
strided data layout and shared memory gives a speedup of 2.2x and 2.88x when compared to non-shared contiguous data for float 
and double precision, respectively. When CUDA experience is limited, contiguous, non-shared data would typically be used, and 
given that the changes required by \frameworkSpace compared to such an implementation are only a few lines of code, the performance 
increase is significant. 

\begin{lstlisting}[style=cpp,float=t,caption=Simplified Euler solver graph creation.,label={lst:euler_graph}]
Tensor<State, 2> in({1, 4}, padding, sizex, sizey);
Tensor<State, 2> out({1, 4}, padding, sizex, sizey);
Tensor<double, 2> wavespeeds({1, 4}, sizex, sizey);
ReductionResult<double> max_wavespeed;

Graph solve_graph;
solve_graph
  .split(set_wavespeeds, in, wavespeeds)
  .then_reduce(wavespeeds, max_wavespeed, MaxReducer)
  .then(set_dt, max_wavespeed)
  .then_split(set_boundary, in)
  .then_split(
    update_state_x, concurrent_padded_access(in), out)
  .then_split(set_boundary, out)
  .then_split(
    update_state_y, concurrent_padded_access(out), in)
  .then(swap_in_out, in, out);
\end{lstlisting}

%=================================================================
% SCALING PERFORMANCE
%=================================================================
\section{Scaling Performance}\label{sec:scaling_results}

Ideally, performance of \frameworkSpace would scale linearly with an increase in computational resources, however, with multiple GPUs additional 
data transfer is required which introduces synchronization points, reducing the effectiveness of the parallelism. The
benchmarks in this section are designed to determine the effectiveness of the framework to hide these costs. As a benchmark, we 
choose a solver for the 2D Euler equations since it requires many of the features of the 
framework described previously and will form the basis of a future application paper.

\begin{figure}[!t]
\centering
\includegraphics[width=.96\linewidth]{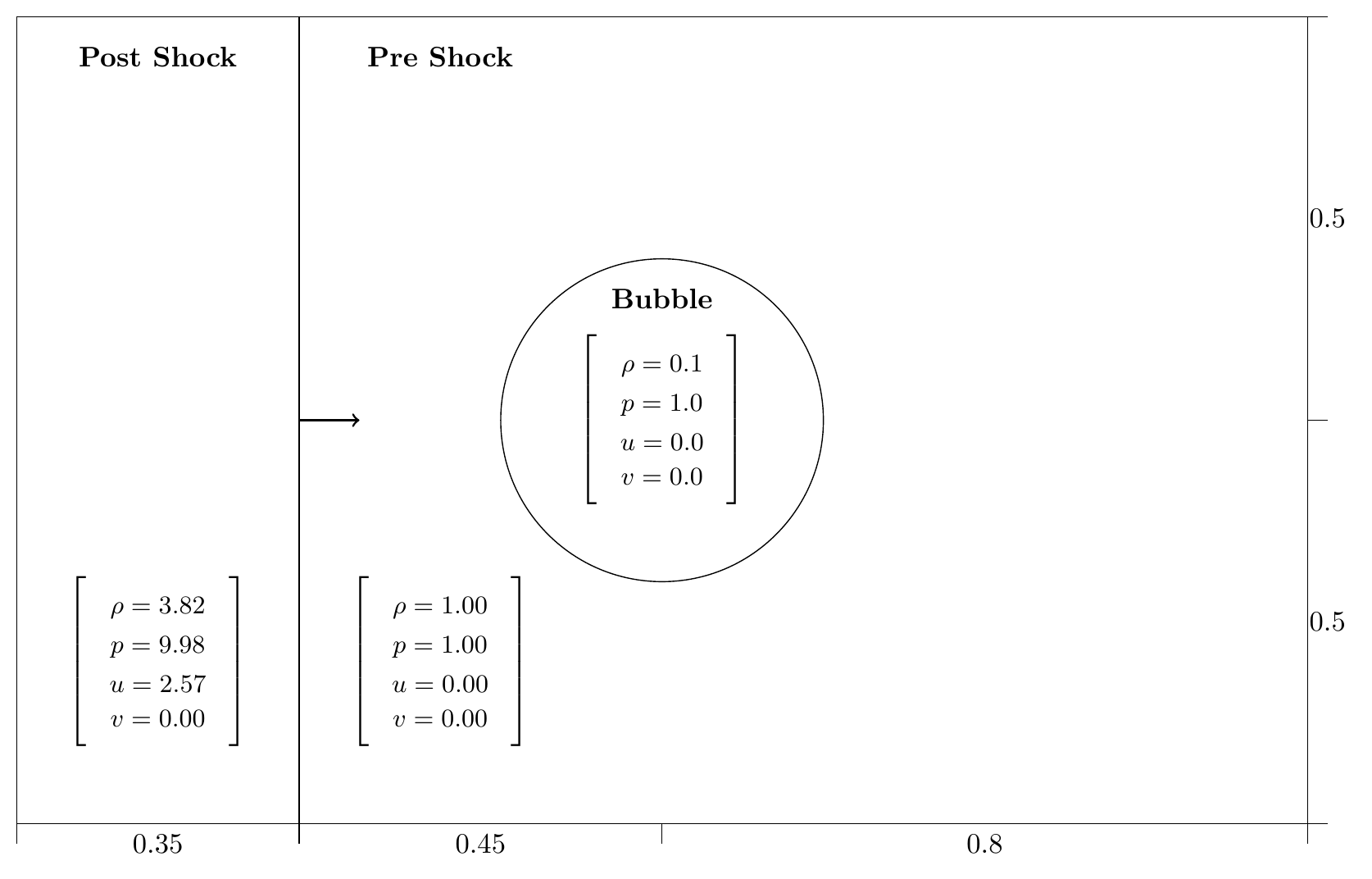}
\caption{Initial conditions for the 2D shock bubble problem.}
\label{fig:shock_bubble}
\end{figure}

The specific problem we choose is a Mach 3.81 shock wave impacting a low density air bubble. This is a problem often used to validate a solver 
in computational fluid dynamics, so provides a good example of how \frameworkSpace can be used to simplify the implementation of 
a real-world problem. A high-level example of the code used to construct the graph which represents the Euler solver 
is shown in \rlist{lst:euler_graph}, while the initial conditions of the problem are shown in \rfig{fig:shock_bubble}. For all 
benchmarks we run the simulation for 1000 time steps.

We also provide the efficiency of \frameworkSpace for both the strong and weak scaling cases, which quantify the percentage overhead 
that is added for each configuration. For weak scaling, the efficiency is computed as
\begin{equation*}
    \textrm{Weak Scaling Efficiency} (\%)= \frac{t_1}{t_N} * 100
\end{equation*}
and for strong scaling as
\begin{equation*}
    \textrm{Strong Scaling Efficiency} (\%) = \frac{t_1}{N * t_N} * 100
\end{equation*}
where for both cases, $t_1$ is the execution time for a single GPU, $N$ is the number of GPUs, and $t_N$ is the execution time for $N$ GPUs.

\subsection{Weak Scaling}

We assign each GPU a partition of the domain, defined by splitting the domain in the $y$ dimension. Each GPU 
requires padding data in the $y$ dimension from either its neighbour(s) or the domain boundary, as well as padding data in the $x$ 
dimension for the domain boundary data. We increase the problem size in both the $x$ and $y$ dimensions when increasing the number of GPUs such 
that the problem configuration is the same relative to the domain. When more GPUs are used, the overall problem size increases, however, each GPU 
always has the same number of cells assigned to it, which to we set to 6.4 million. The results are shown 
in \rfig{fig:weak_scaling}, with the efficiency shown in \rfig{fig:scaling_efficiency}. \frameworkSpace performs well 
in the weak scaling benchmarks, introducing only a small amount of overhead and allowing a very similar computation time for a single GPU compared
to 8 GPUs, as the problem size scales.

\subsection{Strong Scaling}

Again we assign each GPU a partition of the domain and split along the $y$ dimension. For this benchmark we keep the problem size fixed and use two
problem sizes at resolutions of 6400 by 4000 cells (small), for a total of 25.6 million cells, and 9600 by 6000 cells (large), for a total of 57.6 
million cells. Adding GPUs thus decreases the number of cells on which each GPU performs computation. For perfect scaling we would expect an increase in speedup which is linear in the number of GPUs. 
However, since as the number of GPUs increases the relative cost of the overhead for data transfer, 
synchronization, and contention in the work stealing algorithm also increases, linear scaling is not practically achievable.

\begin{figure}[!t]
\centering
\includegraphics[width=.96\linewidth]{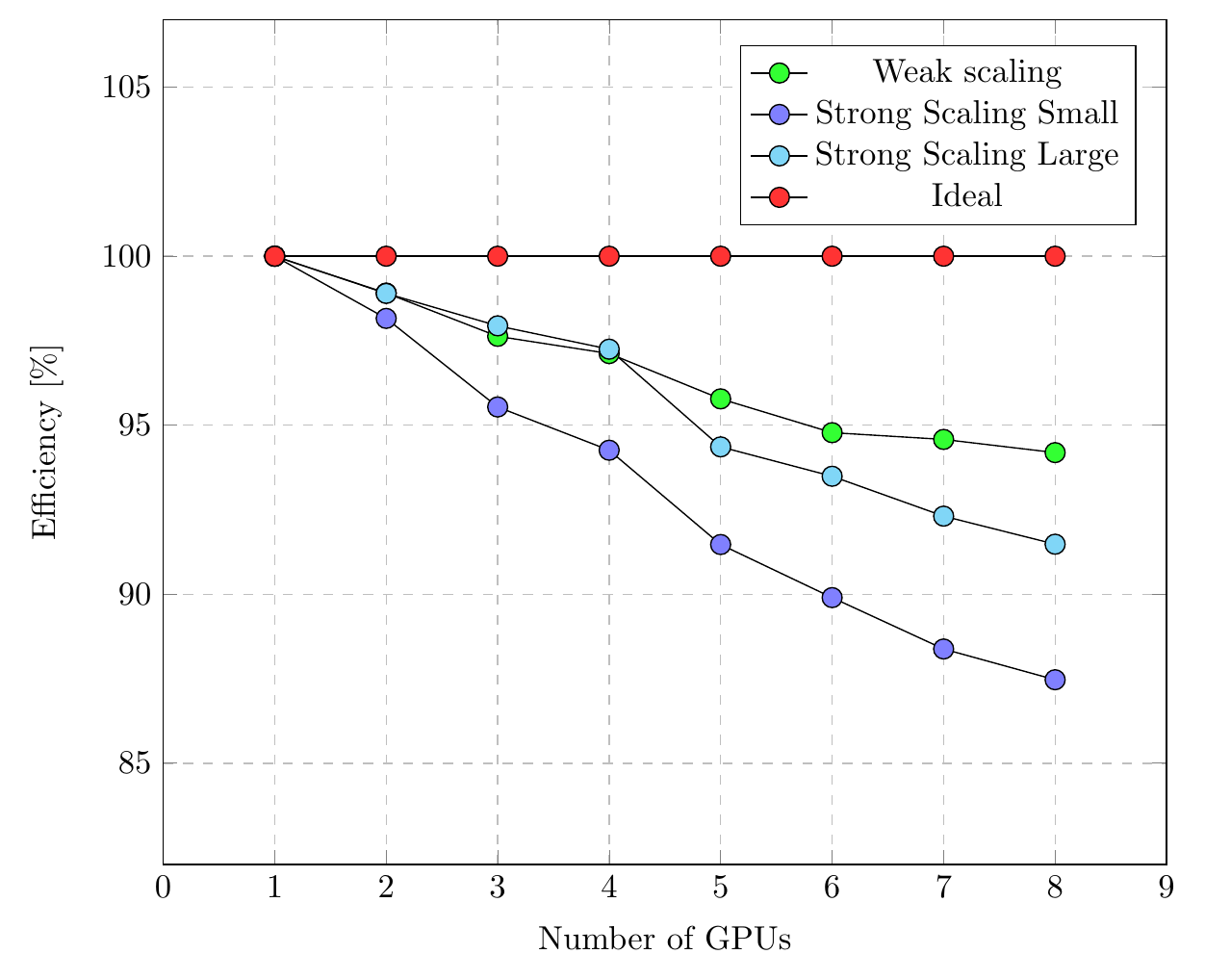}
\caption{Scaling efficiency of \frameworkSpace for strong and weak scaling for a 2D Euler solver.}
\label{fig:scaling_efficiency}
\end{figure}

\begin{figure*}
    \centering
    \begin{subfigure}{.495\linewidth}
    \includegraphics[width=\columnwidth]{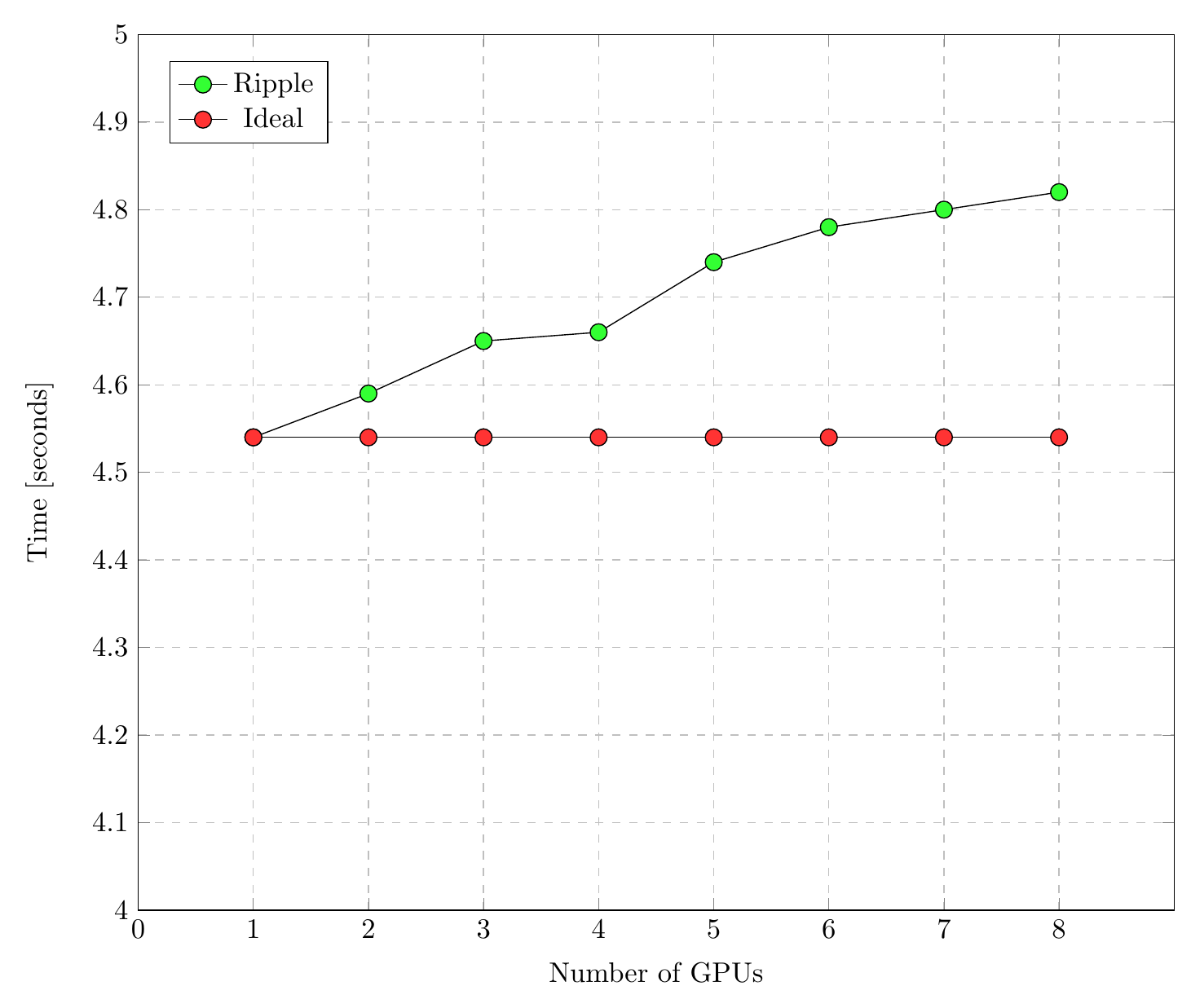}
    \caption{Weak scaling results for \frameworkSpace across multiple GPUs.}
    \label{fig:weak_scaling}
    \end{subfigure}
    \hfill
    \begin{subfigure}{.495\linewidth}
    \includegraphics[width=\columnwidth]{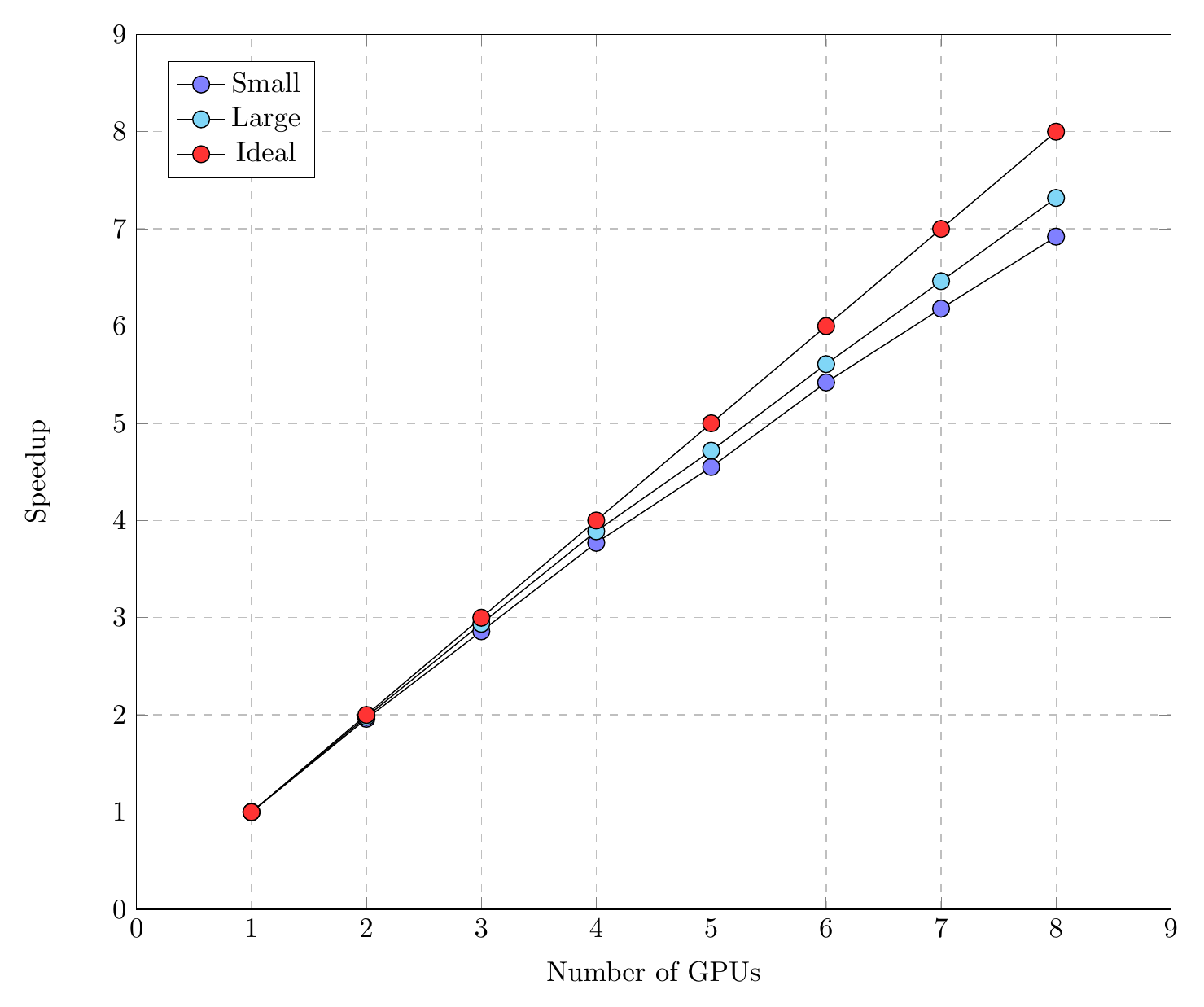}
    \caption{Strong scaling results for \frameworkSpace across multiple GPUs.}
    \label{fig:strong_scaling}
    \end{subfigure}
    \caption{Scaling results for \frameworkSpace for a 2D Euler solver on a shock bubble problem.}
    \label{fig:scaling}
\end{figure*}

The results are shown in \rfig{fig:strong_scaling}, with the efficiency shown in \rfig{fig:scaling_efficiency}. We achieve an increase of 6.92x 
for 8 GPUs at an efficiency of 87.5\% for the small problem, and 7.32x at an efficiency of 91.5\% for the large problem. We see that for the small
problem, the cost of the overhead is high due to the small problem size, while for the larger problem the overhead is minimal since each GPU still
has enough work to better hide communication costs. A significant portion of this reduction comes from 
the NUMA configuration of the node. When the GPUs are connected via NVLINK, the padding data transfers can be performed directly between the 
GPUs without going 
through the CPU (which is the default for \frameworkSpace when possible). However, when the GPUs are split across two sockets, with 4 GPUs per socket, and half the CPU cores on each socket, there is cross socket
communication between the CPUs and GPUs. While the costs of communication between sockets is minimal, it is still present, particularly in the work 
stealing algorithm which may steal tasks resulting in a CPU core submitting work to a GPU which is located on a different socket. These effects can
be seen in all the scaling results, particularly in \rfig{fig:scaling_efficiency} where the scaling between 1-4 and 5-8 GPUs have similar 
gradients, with a significantly larger decrease in efficiency coming from the increase from 4 to 5 GPUs. These results warrant further work into the 
effects of topological based algorithms for work stealing, which may reduce these effects.

%=================================================================
% FUTURE WORK
%=================================================================
\section{Limitations \& Future Work}\label{sec:future_work}

The main limitation of our library is the neighbouring region for which data can be accessed. It is only possible
to access data which is on the same device as the data to which an iterator points. Therefore, if a domain is
partitioned in two, one half of the domain cannot access data from the other half. However, this is true of
any application which uses multiple devices, and the data would have to be sorted so that any elements which need
to access each other are on the same device. Future work would add support for manipulation of the tensor data type
to provide support for these kinds of use cases.

Currently, while very close to providing a unified interface for kernels which can be executed on both the CPU
and the GPU, \frameworkSpace has some limitations in this area. The first is in kernels which require synchronization of threads on 
the GPU, through \verb|__syncthreads()|, or our abstraction \verb|ripple::syncthreads()|. On the CPU, this would
require that a thread executes part of a function until the \verb|syncthreads| call, saves the state for the thread,
does the same for all other threads, and then resumes back at the first thread after restoring the state. This is
challenging in C++; however, with the recent introduction of coroutines, it is possible to suspend a function which
can later be resumed, and one area of future work is to provide this support so that there is a unified interface for
both the CPU and the GPU. This would also be a starting point for allowing kernels which can be split across GPUs and
CPUs, executing simultaneously, when a sufficient number of CPU cores are present.

As mentioned in \rsec{sec:particle_update}, NVIDIA's latest Ampere architecture allows for asynchronous data transfer
between global and shared memory, with a more fine-grained synchronization model. We intend to use this to improve
the performance of our library for shared memory, specifically to reduce the overhead of checking the validity of the
iterators.

Improving the functionality of the allocators and the ability to create hierarchical tensor structures is 
another area of future work. Many applications have regions of interest within a domain which are distributed
sparsely within the global computational domain, requiring adaptive, hierarchical data structures. For optimal
performance on the GPU, we have shown that memory access patterns play a large role, and thus it would be ideal to
allow the same access patterns but have adaptive resolution. Our tensor data structure and iterator abstractions could
be extended to provide such functionality for other data types such as k-d trees, octrees, 
adaptive meshes, and bounding volume hierarchies, for example.

Lastly, our library is designed for single nodes with multiple GPUs and CPUs. While modern compute nodes allow for many
GPUs and hundreds of CPU cores with hundreds of GB of RAM for both the CPU and GPU, some problems require still larger 
computational resources, and thus we could extend our library to allow for computation across any number of nodes.

%=================================================================
% CONCLUSION
%=================================================================
\section{Conclusion}\label{sec:conclusion}

We have presented the abstractions and details of our library, \framework. Our main contributions are 
the ability for user-defined types to have polymorphic layouts to achieve the best possible performance
on GPU accelerators, a multi-dimensional tensor data type and iterator classes over the multi-dimensional
space, a graph interface for specifying the structure of a computation, and an efficient scheduler
for execution of the computational graph. Lastly, our interface allows computations to scale across all
accelerators on a node, allowing high-performance computations over very large domains to be implemented
with minimal programmer effort.

We have used micro benchmarks to illustrate the effectiveness of data layout polymorphism as well as the
importance of utilising different physical memory spaces available in GPU accelerators. We also show
how all these options are problem dependent, and that our library exposes these through simple interfaces
which allow for rapid testing of the different possible configurations.

Results have been presented for strong and weak scaling cases to illustrate the effectiveness of our library on
real world problems, as well as to evaluate the overhead of the abstractions provided by the library and its performance to existing frameworks. For weak
scaling we achieve around 95\% for 8 GPUs, while for strong scaling we achieve performance scaling of between
6.9 and 7.3x for 8 GPUs, for a common problem from computational fluid dynamics.

\frameworkSpace is under constant development and we plan to further investigate additional data structures, scheduling
algorithms, and allowing our tensor data type to increase resolution in sub-regions, dynamically, for applications where
computation is concentrated in certain regions. We also intend to extend the CPU interface to allow CUDA constructs
such as \verb|__syncthreads()| to work in both computational spaces, and lastly to scale to multiple nodes to allow 
even larger problems with minimal knowledge of parallel and distributed systems.



%%       \citet{<label>} ==> Jones et al. [21]
%%       \citep{<label>} ==> [21]
%%

\bibliographystyle{elsarticle-num-names} 
\bibliography{ripple}

\begin{thebibliography}{26}
\expandafter\ifx\csname natexlab\endcsname\relax\def\natexlab#1{#1}\fi
\providecommand{\url}[1]{\texttt{#1}}
\providecommand{\href}[2]{#2}
\providecommand{\path}[1]{#1}
\providecommand{\DOIprefix}{doi:}
\providecommand{\ArXivprefix}{arXiv:}
\providecommand{\URLprefix}{URL: }
\providecommand{\Pubmedprefix}{pmid:}
\providecommand{\doi}[1]{\href{http://dx.doi.org/#1}{\path{#1}}}
\providecommand{\Pubmed}[1]{\href{pmid:#1}{\path{#1}}}
\providecommand{\bibinfo}[2]{#2}
\ifx\xfnm\relax \def\xfnm[#1]{\unskip,\space#1}\fi
%Type = Misc
\bibitem[{{NVIDIA}(2020)}]{Cuda2020}
\bibinfo{author}{{NVIDIA}}, \bibinfo{title}{{CUDA} toolkit documentation},
  \bibinfo{year}{2020}. \URLprefix
  \url{https://docs.nvidia.com/cuda/index.html}.
%Type = Misc
\bibitem[{{The Khronos SYCL Working Group}(2021)}]{sycl}
\bibinfo{author}{{The Khronos SYCL Working Group}}, \bibinfo{title}{Sycl 2020
  specification (revision 2)}, \bibinfo{year}{2021}. \URLprefix
  \url{https://www.khronos.org/registry/SYCL/specs/sycl-2020/pdf/sycl-2020.pdf}.
%Type = Book
\bibitem[{{OpenMP Architecture Review Board}(2018)}]{openmp}
\bibinfo{author}{{OpenMP Architecture Review Board}}, \bibinfo{title}{OpenMP
  Application Programming Interface}, \bibinfo{edition}{5.0} ed.,
  \bibinfo{year}{2018}.
%Type = Book
\bibitem[{OpenACC-Standard.org(2019)}]{openacc}
\bibinfo{author}{OpenACC-Standard.org}, \bibinfo{title}{The OpenACCC
  Application Programming Interface}, \bibinfo{edition}{3.0} ed.,
  \bibinfo{year}{2019}.
%Type = Book
\bibitem[{{BSC Programming Models}(2019)}]{openmps}
\bibinfo{author}{{BSC Programming Models}}, \bibinfo{title}{OmpSs
  Specification}, \bibinfo{year}{2019}.
%Type = Inproceedings
\bibitem[{{Lee} and {Vetter}(2012)}]{Lee2012}
\bibinfo{author}{S.~{Lee}}, \bibinfo{author}{J.~S. {Vetter}},
\newblock \bibinfo{title}{Early evaluation of directive-based {GPU} programming
  models for productive exascale computing},
\newblock in: \bibinfo{booktitle}{SC '12: Proceedings of the International
  Conference on High Performance Computing, Networking, Storage and Analysis},
  \bibinfo{year}{2012}, pp. \bibinfo{pages}{1--11}.
%Type = Article
\bibitem[{Edwards] et~al.(2014)Edwards], Trott, and Sunderland}]{EDWARDS2014}
\bibinfo{author}{H.~C. Edwards]}, \bibinfo{author}{C.~R. Trott},
  \bibinfo{author}{D.~Sunderland},
\newblock \bibinfo{title}{Kokkos: Enabling manycore performance portability
  through polymorphic memory access patterns},
\newblock \bibinfo{journal}{Journal of Parallel and Distributed Computing}
  \bibinfo{volume}{74} (\bibinfo{year}{2014}) \bibinfo{pages}{3202 -- 3216}.
  \DOIprefix\doi{https://doi.org/10.1016/j.jpdc.2014.07.003},
  \bibinfo{note}{domain-Specific Languages and High-Level Frameworks for
  High-Performance Computing}.
%Type = Article
\bibitem[{Augonnet et~al.(2011)Augonnet, Thibault, Namyst, and
  Wacrenier}]{Augonnet2011}
\bibinfo{author}{C.~Augonnet}, \bibinfo{author}{S.~Thibault},
  \bibinfo{author}{R.~Namyst}, \bibinfo{author}{P.-A. Wacrenier},
\newblock \bibinfo{title}{Starpu: a unified platform for task scheduling on
  heterogeneous multicore architectures},
\newblock \bibinfo{journal}{Concurrency and Computation: Practice and
  Experience} \bibinfo{volume}{23} (\bibinfo{year}{2011})
  \bibinfo{pages}{187--198}. \DOIprefix\doi{10.1002/cpe.1631}.
%Type = Inproceedings
\bibitem[{Kaiser et~al.(2014)Kaiser, Heller, Adelstein-Lelbach, Serio, and
  Fey}]{kaiser2014}
\bibinfo{author}{H.~Kaiser}, \bibinfo{author}{T.~Heller},
  \bibinfo{author}{B.~Adelstein-Lelbach}, \bibinfo{author}{A.~Serio},
  \bibinfo{author}{D.~Fey},
\newblock \bibinfo{title}{Hpx: A task based programming model in a global
  address space},
\newblock in: \bibinfo{booktitle}{Proceedings of the 8th International
  Conference on Partitioned Global Address Space Programming Models}, PGAS
  ’14, \bibinfo{publisher}{Association for Computing Machinery},
  \bibinfo{address}{New York, NY, USA}, \bibinfo{year}{2014}.
  \DOIprefix\doi{10.1145/2676870.2676883}.
%Type = Misc
\bibitem[{Huang et~al.(2020)Huang, Lin, Lin, and Lin}]{huang2020}
\bibinfo{author}{T.-W. Huang}, \bibinfo{author}{D.-L. Lin},
  \bibinfo{author}{Y.~Lin}, \bibinfo{author}{C.-X. Lin},
  \bibinfo{title}{Cpp-taskflow v2: A general-purpose parallel and heterogeneous
  task programming system at scale}, \bibinfo{year}{2020}.
  \href{http://arxiv.org/abs/2004.10908}{{\tt arXiv:2004.10908}}.
%Type = Inproceedings
\bibitem[{Belviranli et~al.(2016)Belviranli, Khorasani, Bhuyan, and
  Gupta}]{Belviranli2016}
\bibinfo{author}{M.~E. Belviranli}, \bibinfo{author}{F.~Khorasani},
  \bibinfo{author}{L.~N. Bhuyan}, \bibinfo{author}{R.~Gupta},
\newblock \bibinfo{title}{Cumas: Data transfer aware multi-application
  scheduling for shared gpus},
\newblock in: \bibinfo{booktitle}{Proceedings of the 2016 International
  Conference on Supercomputing}, ICS ’16, \bibinfo{publisher}{Association for
  Computing Machinery}, \bibinfo{address}{New York, NY, USA},
  \bibinfo{year}{2016}. \URLprefix
  \url{https://doi.org/10.1145/2925426.2926271}.
  \DOIprefix\doi{10.1145/2925426.2926271}.
%Type = Inproceedings
\bibitem[{{Bastem} et~al.(2017){Bastem}, {Unat}, {Zhang}, {Almgren}, and
  {Shalf}}]{Bastem2017}
\bibinfo{author}{B.~{Bastem}}, \bibinfo{author}{D.~{Unat}},
  \bibinfo{author}{W.~{Zhang}}, \bibinfo{author}{A.~{Almgren}},
  \bibinfo{author}{J.~{Shalf}},
\newblock \bibinfo{title}{Overlapping data transfers with computation on {GPU}
  with tiles},
\newblock in: \bibinfo{booktitle}{2017 46th International Conference on
  Parallel Processing (ICPP)}, \bibinfo{year}{2017}, pp.
  \bibinfo{pages}{171--180}.
%Type = Inproceedings
\bibitem[{{Wahib} et~al.(2016){Wahib}, {Maruyama}, and {Aoki}}]{Wahib2016}
\bibinfo{author}{M.~{Wahib}}, \bibinfo{author}{N.~{Maruyama}},
  \bibinfo{author}{T.~{Aoki}},
\newblock \bibinfo{title}{Daino: A high-level framework for parallel and
  efficient {AMR} on {GPU}s},
\newblock in: \bibinfo{booktitle}{SC '16: Proceedings of the International
  Conference for High Performance Computing, Networking, Storage and Analysis},
  \bibinfo{year}{2016}, pp. \bibinfo{pages}{621--632}.
%Type = Article
\bibitem[{Dean and Ghemawat(2008)}]{Dean2008}
\bibinfo{author}{J.~Dean}, \bibinfo{author}{S.~Ghemawat},
\newblock \bibinfo{title}{Mapreduce: Simplified data processing on large
  clusters},
\newblock \bibinfo{journal}{Commun. ACM} \bibinfo{volume}{51}
  (\bibinfo{year}{2008}) \bibinfo{pages}{107–113}.
  \DOIprefix\doi{10.1145/1327452.1327492}.
%Type = Article
\bibitem[{Leijen et~al.(2009)Leijen, Schulte, and Burckhardt}]{Leijan2009}
\bibinfo{author}{D.~Leijen}, \bibinfo{author}{W.~Schulte},
  \bibinfo{author}{S.~Burckhardt},
\newblock \bibinfo{title}{The design of a task parallel library},
\newblock \bibinfo{journal}{SIGPLAN Not.} \bibinfo{volume}{44}
  (\bibinfo{year}{2009}) \bibinfo{pages}{227–242}.
  \DOIprefix\doi{10.1145/1639949.1640106}.
%Type = Inproceedings
\bibitem[{{Lee} and {Eigenmann}(2010)}]{Lee2010}
\bibinfo{author}{S.~{Lee}}, \bibinfo{author}{R.~{Eigenmann}},
\newblock \bibinfo{title}{{OpenMPC}: Extended {OpenMP} programming and tuning
  for {GPU}s},
\newblock in: \bibinfo{booktitle}{SC '10: Proceedings of the 2010 ACM/IEEE
  International Conference for High Performance Computing, Networking, Storage
  and Analysis}, \bibinfo{year}{2010}, pp. \bibinfo{pages}{1--11}.
  \DOIprefix\doi{10.1109/SC.2010.36}.
%Type = Article
\bibitem[{Blumofe and Leiserson(1999)}]{Blumofe1999}
\bibinfo{author}{R.~D. Blumofe}, \bibinfo{author}{C.~E. Leiserson},
\newblock \bibinfo{title}{Scheduling multithreaded computations by work
  stealing},
\newblock \bibinfo{journal}{J. ACM} \bibinfo{volume}{46} (\bibinfo{year}{1999})
  \bibinfo{pages}{720–748}. \DOIprefix\doi{10.1145/324133.324234}.
%Type = Inproceedings
\bibitem[{Arora et~al.(1998)Arora, Blumofe, and Plaxton}]{Arora1998}
\bibinfo{author}{N.~S. Arora}, \bibinfo{author}{R.~D. Blumofe},
  \bibinfo{author}{C.~G. Plaxton},
\newblock \bibinfo{title}{Thread scheduling for multiprogrammed
  multiprocessors},
\newblock in: \bibinfo{booktitle}{Proceedings of the Tenth Annual ACM Symposium
  on Parallel Algorithms and Architectures}, SPAA '98,
  \bibinfo{publisher}{Association for Computing Machinery},
  \bibinfo{address}{New York, NY, USA}, \bibinfo{year}{1998}, p.
  \bibinfo{pages}{119–129}. \DOIprefix\doi{10.1145/277651.277678}.
%Type = Inproceedings
\bibitem[{{Agrawal} et~al.(2010){Agrawal}, {Leiserson}, and
  {Sukha}}]{Agrawal2010}
\bibinfo{author}{K.~{Agrawal}}, \bibinfo{author}{C.~E. {Leiserson}},
  \bibinfo{author}{J.~{Sukha}},
\newblock \bibinfo{title}{Executing task graphs using work-stealing},
\newblock in: \bibinfo{booktitle}{2010 IEEE International Symposium on Parallel
  Distributed Processing (IPDPS)}, \bibinfo{year}{2010}, pp.
  \bibinfo{pages}{1--12}. \DOIprefix\doi{10.1109/IPDPS.2010.5470403}.
%Type = Inproceedings
\bibitem[{Chase and Lev(2005)}]{Chase2005}
\bibinfo{author}{D.~Chase}, \bibinfo{author}{Y.~Lev},
\newblock \bibinfo{title}{Dynamic circular work-stealing deque},
\newblock in: \bibinfo{booktitle}{Proceedings of the Seventeenth Annual ACM
  Symposium on Parallelism in Algorithms and Architectures}, SPAA '05,
  \bibinfo{publisher}{Association for Computing Machinery},
  \bibinfo{address}{New York, NY, USA}, \bibinfo{year}{2005}, p.
  \bibinfo{pages}{21–28}. \DOIprefix\doi{10.1145/1073970.1073974}.
%Type = Article
\bibitem[{L\^{e} et~al.(2013)L\^{e}, Pop, Cohen, and Zappa~Nardelli}]{Le2013}
\bibinfo{author}{N.~M. L\^{e}}, \bibinfo{author}{A.~Pop},
  \bibinfo{author}{A.~Cohen}, \bibinfo{author}{F.~Zappa~Nardelli},
\newblock \bibinfo{title}{Correct and efficient work-stealing for weak memory
  models},
\newblock \bibinfo{journal}{SIGPLAN Not.} \bibinfo{volume}{48}
  (\bibinfo{year}{2013}) \bibinfo{pages}{69–80}.
  \DOIprefix\doi{10.1145/2517327.2442524}.
%Type = Inproceedings
\bibitem[{Toro(1996)}]{Toro1996}
\bibinfo{author}{E.~F. Toro},
\newblock \bibinfo{title}{On {G}limm-related schemes for conservation laws},
\newblock \bibinfo{year}{1996}.
%Type = Article
\bibitem[{Sethian(1996)}]{Sethian1996}
\bibinfo{author}{J.~A. Sethian},
\newblock \bibinfo{title}{A fast marching level set method for monotonically
  advancing fronts},
\newblock \bibinfo{journal}{Proceedings of the National Academy of Sciences}
  \bibinfo{volume}{93} (\bibinfo{year}{1996}) \bibinfo{pages}{1591--1595}.
  \DOIprefix\doi{10.1073/pnas.93.4.1591}.
  \href{http://arxiv.org/abs/https://www.pnas.org/content/93/4/1591.full.pdf}{{\tt
  arXiv:https://www.pnas.org/content/93/4/1591.full.pdf}}.
%Type = Article
\bibitem[{Zhao(2005)}]{Zhao2005}
\bibinfo{author}{H.~Zhao},
\newblock \bibinfo{title}{A fast sweeping method for eikonal equations},
\newblock \bibinfo{journal}{Mathematics of computation}
  (\bibinfo{year}{2005}).
%Type = Article
\bibitem[{Zhao(2006)}]{Zhao2006}
\bibinfo{author}{H.~Zhao},
\newblock \bibinfo{title}{Parallel implementation of the fast sweeping method},
\newblock \bibinfo{journal}{International Journal of Computer Mathematics -
  IJCM} \bibinfo{volume}{25} (\bibinfo{year}{2006}).
%Type = Article
\bibitem[{Jeong and Whitaker(2008)}]{Jeong2008}
\bibinfo{author}{W.~Jeong}, \bibinfo{author}{R.~Whitaker},
\newblock \bibinfo{title}{A fast iterative method for eikonal equations},
\newblock \bibinfo{journal}{SIAM J. Sci. Comput.} \bibinfo{volume}{30}
  (\bibinfo{year}{2008}) \bibinfo{pages}{2512--2534}.

\end{thebibliography}


%\begin{thebibliography}{00}



%\end{thebibliography}
\end{document}